\documentclass[journal]{IEEEtran}
\ifCLASSINFOpdf
\else
\fi
\usepackage[portuguese,english]{babel}
\usepackage[utf8]{inputenc} 
\usepackage[T1]{fontenc}
\usepackage{url} 
\usepackage{hyperref}
\usepackage{enumerate}
\usepackage{booktabs}
\usepackage{pgfplots}
\usepackage{cite}
\usepackage{xcolor,colortbl}
\usepackage{bm}
\usepackage{upgreek}
\usepackage{amsmath}
\usepackage{caption}
\usepackage{subcaption}
\usepackage{graphicx}
\usepackage{tabstackengine}
\usepackage{amsfonts} 
\usepackage{mathtools, nccmath}
\usepackage{environ}         
\usepackage{etoolbox}        
\usepackage{graphicx}        
\usepackage{algorithm}
\usepackage{gensymb}

\newlength{\myl}
\let\origequation=\equation
\let\origendequation=\endequation

\RenewEnviron{equation}{
  \settowidth{\myl}{$\BODY$}                       
  \origequation
  \ifdimcomp{\the\linewidth}{>}{\the\myl}
  {\ensuremath{\BODY}}                             
  {\resizebox{0.95\linewidth}{!}{\ensuremath{\BODY}}}  
  \origendequation
}
\ExplSyntaxOn
\NewDocumentCommand\Vector{m}
 {
  \commexo_vector:n { #1 }
 }

\cs_new_protected:Npn \commexo_vector:n #1
 {
  \tl_map_inline:nn { #1 }
   {
    \commexo_vector_inner:n { ##1 }
   }
 }

\cs_new_protected:Npn \commexo_vector_inner:n #1
 {
  \tl_if_in:VnTF \g_commexo_latin_tl { #1 }
   {
    \mathbf { #1 } 
   }
   {
    \tl_if_in:VnTF \g_commexo_ucgreek_tl { #1 }
     {
      \bm { #1 } 
     }
     {
      \tl_if_in:VnTF \g_commexo_lcgreek_tl { #1 }
       {
        \commexo_makeboldupright:n { #1 }
       }
       {
        #1 
       }
     }
   }
 }

\cs_new_protected:Npn \commexo_makeboldupright:n #1
 {
  \bm { \use:c { up \cs_to_str:N #1 } }
 }

\tl_new:N \g_commexo_latin_tl
\tl_new:N \g_commexo_ucgreek_tl
\tl_new:N \g_commexo_lcgreek_tl
\tl_gset:Nn \g_commexo_latin_tl
 {
  ABCDEFGHIJKLMNOPQRSTUVWXYZ
  abcdefghijklmnopqrstuvwxyz
 }
\tl_gset:Nn \g_commexo_ucgreek_tl
 {
  \Gamma\Delta\Theta\Lambda\Pi\Sigma\Upsilon\Phi\Chi\Psi\Omega
 }
\tl_gset:Nn \g_commexo_lcgreek_tl
 {
  \alpha\beta\gamma\delta\epsilon\zeta\eta\theta\iota\kappa
  \lambda\mu\nu\xi\pi\rho\sigma\tau\upsilon\phi\chi\psi\omega
  \varepsilon\vartheta\varpi\varphi\varsigma\varrho
 }

\ExplSyntaxOff

\setstackEOL{;}
\setstackTAB{,}
\setstacktabbedgap{1ex}
\setstackgap{L}{2.2\normalbaselineskip}
\let\nmatrix\bracketMatrixstack

\newcommand{\vc}[1]{\Vector{#1}}

\newcommand{\set}[1]{\mathcal{#1}}
\newcommand{\abs}[1]{\left|#1\right|}
\newcommand{\floor}[1]{\left\lfloor #1 \right\rfloor}
\newcommand{\lfc}{\mathcal{L}}
\newcommand{\lf}{\vc{L}}

\newcommand{\T}{^{\rm T}}

\definecolor{Gray}{gray}{0.85}
\DeclarePairedDelimiter{\nint}{\lfloor}{\rceil} 

 \pgfplotsset{compat=1.18}

\begin{document}
%
\title{Iterative Occlusion-Aware Light Field Depth Estimation using 4D Geometrical Cues }
%
%
%

\author{Rui~Lourenco$^{1,3}$,~\IEEEmembership{Student Member,~IEEE,}
        Lucas~Thomaz$^{1,2}$,~\IEEEmembership{Senior Member,~IEEE,}
        Eduardo A. B. Silva$^{3}$,~\IEEEmembership{Senior Member,~IEEE,} 
        and~Sergio~M.~M.~Faria$^{1,2}$,~\IEEEmembership{Senior Member,~IEEE}
\thanks{$1$ Instituto de Telecomunicações, Portugal}
\thanks{2 ESTG - Polytechnic University of Leiria, Leiria, Portugal}
\thanks{3 PEE, COPPE, Federal University of Rio de Janeiro, Rio de Janeiro, Brazil}%
\thanks{This work was supported by the Funda\c{c}\~ao para a Ci\^encia e a Tecnologia (FCT), Portugal under projects 2023.07886.CEECIND (DOI:10.54499/2023.07886.CEECIND/CP2862/CT0003), Programa Operacional Regional do Centro, and by FCT/MCTES through national funds and when applicable co-funded by EU funds under the project UIDB/EEA/50008/2020 (DOI: 10.54499/UIDB/50008/2020) and LA/P/0109/2020 (DOI: 10.54499/LA/P/0109/2020).}}

%
%

\markboth{Journal of \LaTeX\ Class Files,~Vol.~14, No.~8, August~2015}%
{Shell \MakeLowercase{\textit{et al.}}: Bare Demo of IEEEtran.cls for IEEE Journals}
%



\maketitle

\begin{abstract}

Light field cameras and multi-camera arrays have emerged as promising solutions for accurately estimating depth by passively capturing light information. This is possible because the 3D information of a scene is embedded in the 4D light field geometry. Commonly, depth estimation methods extract this information relying on gradient information, heuristic-based optimisation models, or learning-based approaches. This paper focuses mainly on explicitly understanding and exploiting 4D geometrical cues for light field depth estimation. Thus, a novel method is proposed, based on a non-learning-based optimisation approach for depth estimation that explicitly considers surface normal accuracy and occlusion regions by utilising a fully explainable 4D geometric model of the light field. The 4D model performs depth/disparity estimation by determining the orientations and analysing the intersections of key 2D planes in 4D space, which are the images of 3D-space points in the 4D light field. Experimental results show that the proposed method outperforms both learning-based and non-learning-based state-of-the-art methods in terms of surface normal angle accuracy, achieving a Median Angle Error on planar surfaces, on average, 26.3\% lower than the state-of-the-art, and still being competitive with state-of-the-art methods in terms of MSE $\vc{\times}$ 100 and Badpix 0.07.
\end{abstract}


%
\IEEEpeerreviewmaketitle

\section{Introduction}
The explosion of public and academic interest in Augmented and Virtual Reality applications in recent years~\cite{Cipresso2018,Dargan2023} has prompted the development of advanced imaging techniques to enhance the immersive experience. Among these techniques, light field cameras and multi-camera arrays have gained significant attention due to their ability to capture rich spatial and angular information about a scene. By recording the light rays from multiple viewpoints, these devices enable several applications, from the construction of new points-of-view for a given scene and refocusing of an image to the estimation of the depth of a scene, enabling 3D reconstruction applications. Most importantly, the dense capture of information is used in several computer vision applications, such as automatic measurements and quality control in different types of industries~\cite{raytrix}, post-processing effects on photographs~\cite{Liu2015}, and even the diagnosis of severe medical conditions, such as skin cancer~\cite{Pereira2022}. 

Light field disparity estimation is crucial in many typical applications of light field technology. Unlike other depth estimation technologies, such as structured light~\cite{Scharstein2003} and Light Detection and Ranging (LiDAR)~\cite{Wang2021} systems, light field disparity estimation does not struggle in strong lighting conditions as it does not rely on active sensors. Furthermore, as light fields commonly have a narrow baseline, light field-based methods can overcome the limitations of traditional stereo-vision approaches, increasing accuracy.

The best-performing state-of-the-art (SOTA) methods for light field disparity estimation primarily rely on supervised learning models, such as~\cite{Shin2018,Tsai2020,Kunyuan2020,Yan2022,Wang2022,Han2023,Wentao2023}. These methods provide highly accurate results for the available computer-generated light field datasets, obtaining very good results in terms of most objective accuracy metrics, such as the Mean Squared Error (MSE) or Badpix 0.07, 
as defined in~\cite{Johansen2017}.  

{Unsupervised learning-based methods, such as ~\cite{Peng2018,Zhou2020,Jin2022,Zhang2023,Zhou2024,xiao2025}, provide alternatives to supervised learning models that do not rely on datasets with ground truth depth results. However, the best unsupervised learning-based methods still fall behind supervised-learning-based methods and the best non-learning-based methods.}

Other SOTA methods tend to narrow the focus to 2D cuts of the entire 4D light field, referred to as Epipolar Plane Images (EPIs)~\cite{Dansereau2004,Wanner2012,Li2015,Lourenco2018,Lourenco2022,khan2021,Lee2021,Zhang2016,Schilling2018}, or operate based on energy cost models that avoid some of the known limitations for light field disparity estimation through different sets of heuristics~\cite{Tao2013,Jeon2015,Lin2015,Wang16,Williem16,Strecke2017,Williem2018,Ma2021,Kang2021}. Whilst some of these methods obtain competitive results, they tend to fall behind learning-based methods in terms of objective accuracy metrics.

While the above methods tend to improve depth estimation performance relative to their predecessors, they, in general, do not attempt to build a cohesive mathematical model that fully exploits the four-dimensional complexity of the 4D light field. In this context, the main motivation of the proposed article is to build a framework for exploiting the geometric relations between 4D light field space and 3D scene space. To this end, the article provides a formal geometrical description of light field disparity estimation by introducing the concept of the 4D-Point Projection Plane (4D-PPP), which is the image of a 3D-space point in the 4D light field. 


{ To assess the validity and usefulness of the geometric foundations underlying the proposed framework, a robust occlusion-aware energy cost model is built based on it. In addition, the Iterative Occlusion-Aware Depth Refinement (IOADR) algorithm, which uses novel geometrically informed heuristics to optimise the proposed energy cost, is proposed. 

IOADR proves to be competitive with SOTA unsupervised methods in terms of MSE $\times$ 100 and badpix 0.07, only falling behind supervised methods. In addition, it outperforms both supervised and unsupervised SOTA methods in terms of the Median Angle Error (MAE) in Planar Regions metric, proposed in~ \cite{Johansen2017}. These results provide evidence that the proposed geometric foundation is valuable and worthy pursuing in light field disparity estimation research.}


The remainder of this paper is organised as follows: Section \ref{sec:backgroud} provides a background of related work in light field disparity estimation, highlighting the merits and limitations of existing approaches. {Section \ref{sec:contribution-walktrhough} provides a brief walkthrough of the different contributions presented in this article, and Section \ref{sec:4dppp} presents a formal mathematical description of 4D light field geometry, how depth information is embedded in this geometry and how it can be retrieved. Section \ref{sec:cost-model} presents the proposed novel energy cost model, and Section \ref{sec:ioadi} introduces a novel cost minimisation algorithm for depth map refinement.} Section \ref{sec:results} presents a comparative experimental evaluation regarding the SOTA and ablation studies. Finally, Section \ref{sec:conclusion} concludes the article and outlines potential avenues for future research.

\section{Background}
\label{sec:backgroud}

{ In this work, a light field is regarded as a 4D function that associates to each sample $(u,v)$ from a view located at coordinates $(s,t)$, a vector of colour components $\vc{c} \in \mathbb{R}^3$, such that:  
\begin{equation}
         \vc{c}=\lfc(\vc{u}, \vc{s}),
    \label{eq:continuous-lf}
\end{equation}
where, for $\set{U},\set{S} \subset \mathbb{R}^2$, $\vc{u} = \nmatrix{u,v}\T\in\set{U}$ is referred to as the \textit{spatial position} inside a view, and the view location $\vc{s} = \nmatrix{s,t}^{\rm T}\in\set{S}$ is referred to as the \textit{angular position}.}

Epipolar Plane Images (EPIs) are 2D slices of the light field along either the $s\times u$ or $t\times v$ planes.

{The EPIs reveal an important property of light fields: the image of a 3D-space point $\vc{x}$ in an EPI is a slanted straight line whose slope depends on the depth of point $\vc{x}$~\cite{Carvalho2023}}.


Such properties have been used extensively in the literature to estimate the depths of a 3D scene from light fields. 
In what follows, three classes of depth estimation methods are highlighted: Gradient-based~\cite{Dansereau2004,Wanner2012,Li2015,Lourenco2018,Lourenco2022,khan2021,Lee2021}, energy-model-based~\cite{Tao2013,Jeon2015,Lin2015,Zhang2016,Wang16,Williem16,Strecke2017,Williem2018,Ma2021,Kang2021}, and supervised-learning-based~\cite{Shin2018,Tsai2020,Kunyuan2020}. 

\vspace*{-0.2cm}

\subsection{Gradient-based methods}

Gradient-based methods work by directly estimating the gradient of the geometric structures present in EPIs. This strategy permits depth estimation over a continuous range by determining the angular coefficients of slanted lines in EPIs. However, unless supplemented by post-processing or further optimisation steps, they tend to achieve low accuracy in occluded regions.

An early approach to light field disparity estimation was to directly compute the gradient of EPIs~\cite{Dansereau2004}. More robust approaches make use of the Structure Tensor~\cite{Bigun1987} as a tool that not only measures the direction of the slanted lines in EPIs but also provides a reliability metric for this calculation.

 Wanner et al. \cite{Wanner2012} improve the structure tensor accuracy by calculating disparity using both horizontal and vertical EPIs. Rudin et al.~\cite{Rudin92} proposed a fast Total-Variation-Denoising-based scheme and a global optimisation process. Li \textit{et al.}~\cite{Li2015}, improved this scheme by introducing a penalty metric that weights the reliability measure in a way that improves performance in occlusion regions. Lourenço \textit{et al.}~\cite{Lourenco2022} further improved this paradigm by explicitly comparing the disparity and texture edge maps, in-painting the disparity map with corrected values when a mismatch is found. 

While such methods provide sizeable improvements relative to the base structure tensor, most post-processing improvements and optimisations lack robustness, leading to enlarged silhouettes or introducing algorithmic artefacts.

\subsection{Energy-model-based methods}
Energy-model-based methods create an energy model based on a cost function that should be minimal when the correct depth value is chosen. This minimisation is usually done by building a 3D cost-volume that consists of the costs, according to the energy model, of all combinations of pixel coordinates in a view and a finite set of different disparity labels. Obtaining a disparity map for a view is as simple as finding, for each pixel, the disparity label that minimises this cost. 

Several cost metrics have been introduced based on the constraints of cost-volume minimisation. One of the earliest models, proposed by Tao \textit{et al.} \cite{Tao2013}, combined two different metrics, defocus and correspondence, to provide somewhat accurate results. Lin \textit{et al.}~\cite{Lin2015} improved this approach by refining the energy model. Jeon \textit{et al.}~\cite{Jeon2015} used Fourier analysis and a phase-shift system to build a cost-volume with sub-pixel accuracy. However, all of these methods have issues in the presence of occlusion regions. 

Wang \textit{et al.}~\cite{Wang16} directly improved on~\cite{Tao2013} by relying on edge estimation to model occlusions explicitly. Strecke \textit{et al.}~\cite{Strecke2017} improved on~\cite{Lin2015} by both altering the model to be better behaved in occlusion regions and introducing a joint depth and normal map regularisation. Zhang \textit{et al.} \cite{Zhang2016} proposed the Spinning Parallelogram Operator (SPO), which extends the simple compass operator~\cite{Ruzon99} --- an edge detection and characterisation algorithm --- to the EPI domain, obtaining encouraging results even in occlusion regions.

Williem \textit{et al.}\cite{Williem2018} introduced an entropy-based data cost resilient to occlusions, whereas Kang~\textit{et al.}~\cite{Kang2021} proposed an occlusion-aware voting cost that models occlusions by detecting colour inconsistencies in angular patches. Schilling \textit{et al.} \cite{Schilling2018} achieved notable results by foregoing the cost-volume and, instead, following a local optimisation framework that supports more complex occlusion models, which take into account the depth of nearby pixels.

\subsection{Learning-based methods} \label{subsec:learning-based}
{  More recently, learning-based methods have gained popularity for depth estimation. In general, these works rely on the 4D geometric properties of light fields to adapt existing machine-learning frameworks to the task of estimating depth from light fields.  

Learning-based approaches can be divided into two groups: Supervised and Unsupervised, based on whether or not they rely on ground-truth results in the training step.

\subsubsection{Unsupervised Methods}

Unsupervised learning-based methods provide, on average, lower computational complexity than traditional methods while achieving similar results without requiring the large amounts of ground truth data that supervised methods do.      

The first unsupervised learning-based light field depth-estimation method was proposed by Peng \textit{et al.} \cite{Peng2018}, as a Convolutional Neural Network (CNN) that enforced compliance and divergence constraints on sub-aperture images of the light field. Zhou \textit{et al.}~\cite{Zhou2020} improved on this work by proposing a symmetry loss to handle occlusion areas.  Jin \textit{et al.}\cite{Jin2022} and Zhang \textit{et al.}~\cite{Zhang2023} iterate on Zhou's work by providing different neural networks that explicitly address the occlusion problem. 

Unsupervised networks have broadly achieved the goal of greatly reducing computational complexity. For example, Jin \textit{et al.}'s \cite{Jin2022} proposed algorithm achieves a running time of less than a second in its GPU implementation, which is significantly faster than existing implementations of accurate energy-model-based methods. However, their performance in terms of accuracy still falls behind the ones of both energy-model-based methods and supervised learning-based methods, as shown in Section~\ref{sec:results}.

\subsubsection{Supervised Methods}


In general, supervised learning-based methods present some of the best results known to date in terms of estimation accuracy.
Shin \textit{et al.}~\cite{Shin2018} proposed EPInet, a fully Convolutional Neural Network (CNN) built using a multi-stream network design where each stream receives views with a consistent baseline. Tsai \textit{et al.}  proposed AttNet~\cite{Tsai2020}, which consists of a convolutional neural network with an attention module, while Yan~\textit{et al.}~\cite{Yan2022} improved on this architecture by using light field edges as guidance. Kunyan \textit{et al.}~\cite{Kunyuan2020} present an end-to-end fully convolutional network developed explicitly to estimate the depth value from the orientation of lines on EPIs, taking into account the coherence of relations between such lines.  { Wang \textit{et al.}~\cite{Wang2022} introduce an occlusion aware cost constructor.}  Han \textit{et al.}~\cite{Han2023} extracts the sequential features of EPIs by substituting CNNs with Recursive Neural Networks. { Chao \textit{et al.}~\cite{Wentao2023} introduces sub-pixel disparity learning to a deep neural network. The best supervised learning-based methods outperform energy-model-based methods in terms of accuracy while being more computationally efficient. }

 {  It is important to note that, while learning-based approaches are an active and interesting area of research, the methods based on the exploration of 4D light field geometry proposed in this paper have been developed within the framework of traditional energy-based models. This is so because, although most unsupervised and supervised learning-based methods use concepts derived from 4D light field geometry~\cite{Tsai2020,Yan2022,Han2023}, these geometric concepts are more easily investigated in the context of traditional methods, which provide a greater degree of explainability and provide a straightforward environment to test the viability of the theoretical considerations based on 4D light field geometry introduced in this paper.}}

\section{Proposed Contributions Walkthrough}
{

\label{sec:contribution-walktrhough}

The main framework proposed in this article is explained in three different sections:
\begin{enumerate}
    \item Section~\ref{sec:4dppp}, \textit{\textbf{Depth from 4D light field geometry}},  introduces a formal mathematical description of the 4D light field. The \textit{4D Point-Projection Plane} (4D-PPP) --- a surface in the light field that directly corresponds to a single real-world point --- is parametrised so that \textit{depth} corresponds to the 4D-space orientation of a 4D-PPP. A straightforward method for the estimation of a 4D-PPP orientation is introduced, and limitations of that simple approach to depth estimation are addressed.
    \item  Section~\ref{sec:cost-model},  \textit{\textbf{The Proposed Cost Model}}, addresses the limitations of the simple depth estimation approach presented in Section~\ref{sec:4dppp} by introducing a cost function with three terms: an occlusion-aware data cost term, a colour-orientation congruency term, and a plane geometry term.  
    \item Section~\ref{sec:ioadi}, \textit{\textbf{Iterative Occlusion-Aware Depth Refinement}}, introduces a novel algorithm to minimise the cost function proposed in Secion~\ref{sec:cost-model}. It improves on an initial 4D-PPP orientation map by minimizing the cost model presented in Section \ref{sec:cost-model}. This is performed iteratively by testing several candidate orientations for each pixel and making a robust decision update based on the cost they incur.
\end{enumerate}
}


\section{Depth from 4D Light Field Geometry}
\label{sec:4dppp}

 {
 Obtaining a 3D representation of the captured scene is one of the goals of light field analysis. In this Section, the image of a 3D-space point in the 4D light field is shown to be a 2D plane in 4D space designated as 4D-PPP, and the task of estimating depth from a 4D light Field is reduced to the task of correctly parametrising a 4D-PPP. Lastly, the main limitations of this approach are outlined.
}

{\subsection{The 4D Point Projection Plane}}
\label{subsec:4d-ppp}

The general relationship between the angular and spatial coordinates of a general 4D light field 
is given by~\cite{Johansen2017}:

\begin{equation}
		\vc{u} = D\left(\dfrac{1}{Z_{\rm p}}-\dfrac{1}{z}\right)\vc{s} + D\dfrac{\vc{\xi}^{(x\times y)}(\vc{x})}{z},
 \label{eq:4dppp}
\end{equation}
where  $\vc{u}$ and $\vc{s}$ are light field coordinates as defined in Eq. \eqref{eq:continuous-lf}, $Z_{\rm p}$ is the depth in which the disparity is zero, 
and the 2D vector $\vc{\xi}^{(x\times y)}(\vc{x}) = \nmatrix{x,y}\T$ is the projection of the 3D-space point $\vc{x} = \nmatrix{x,y,z}\T$ on the $x\times y$ plane. Geometrically, the equation describes a 2D-plane in 4D space. This plane is designated as 4D-PPP. 

\begin{figure}
    \centering
    \includegraphics[width=1\linewidth]{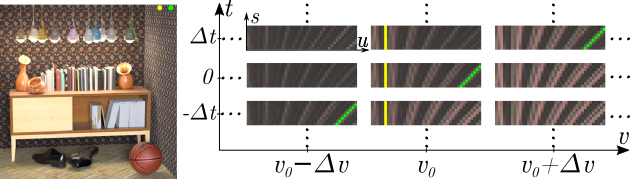}
    \caption{ EPI grid from the \textit{sideboard} light field illustrating two 4D-PPPs. The yellow and green lines are samples of the 4D-PPPs corresponding to the yellow and green 3D scene points highlighted in the centre-view on the left, respectively.}
    \label{fig:epi-grid}
\end{figure}

 An illustration of the 4D-PPP concept is shown in Figure \ref{fig:epi-grid}. There, a 3$\times$3 grid of $su$ slices (EPIs) of the \textit{sideboard}~\cite{Johansen2017} light field, i.e., light field samples for $v = v_0\pm\Delta_v$, $t = t_0\pm\Delta_t$, is shown. 
The yellow and green lines in each EPI are images of the yellow and green 3D scene points shown in the central view on the left. Therefore, these yellow and green lines are samples, on the given 3$\times$3 grid of $su$ slices,  of the 4D-PPPs corresponding to the yellow and green 3D scene points, respectively. 

In accordance to Equation\eqref{eq:4dppp}, the orientations of each of these lines in the $s\times u$ and $t\times v$ EPIs are equal, and are given by $\theta$ in the equation below:
\begin{equation}
    \tan\theta = D\left(\frac{1}{Z_p}-\frac{1}{z}\right).
    \label{eq:theta}
\end{equation}

\subsubsection{4D Point-Projection Planes in Discrete Light Fields}

the light fields used in practice are sampled versions of the continuous light fields.  In this work, a discrete light field $\lf(\vc{m},\vc{k})$ is derived from Eq.~\eqref{eq:continuous-lf} as
\begin{equation}
        \lf(\vc{m},\vc{k}) = \lfc \bigg(\Delta \vc{u}\odot \left(\vc{m}-\vc{m}_{\rm r}\right), \Delta \vc{s} \odot\left(\vc{k}-\vc{k}_{\rm r}\right)\bigg),
        \label{eq:discrete-lf}
\end{equation}
{where $\Delta \vc{s} = \nmatrix{\Delta s, \Delta t}\T\in\mathbb{R}^{2}$ provides the horizontal and vertical baselines, $\Delta \vc{u} = \nmatrix{\Delta u, \Delta v}\T\in\mathbb{R}^{2}$ provides the metric distance between pixels of each view (dot pitches), the operator $\odot$ is the element-wise Hadamard product, $\vc{m} = \nmatrix{m,n}\T\in\set{M}\cup\mathbb{Z}^2$, $\vc{k} = \nmatrix{k,l}\T\in\set{K}\cup\mathbb{Z}^2$, with
\begin{eqnarray}
    \set{M} &=& \{(m,n)\  \big|\  0 \leq m \leq M - 1 ,\ 0 \leq n \leq N - 1 \}, \nonumber \\
    \set{K} &=& \{(k,l)\  \big|\ 0 \leq k \leq K -1,\ 0 \leq l \leq L - 1 \}, 
    \label{eq:lf-boundaries}
\end{eqnarray}
}and $\vc{m}_{\rm r} = \nmatrix{m_{\rm r},n_{\rm r}}\T\in\set{M}\cup\mathbb{Z}^2$ indicates the horizontal and vertical indexes of the origin of the views, and $\vc{k}_{\rm r} = \nmatrix{k_{\rm r}, l_{\rm r}}\T\in\set{K}\cup\mathbb{Z}^2$ indicates the horizontal and vertical indexes of a reference view.
 From Equations \eqref{eq:continuous-lf} and \eqref{eq:discrete-lf}, a discrete light field sample $\vc{p} $ can be computed from a continuous light field position $\nmatrix{\vc{u}{\T},\vc{s}{\T}}{\T}$ using:
    \begin{equation}
		\vc{p}=\nmatrix{\vc{m};\vc{k}} = 
  \nmatrix{
			\nint*{\vc{u}\oslash\Delta \vc{u}+\vc{m}_{\rm r}};
			\nint*{\vc{s}\oslash\Delta \vc{s}+\vc{k}_{\rm r}}
		},
        \label{eq:discrete-coordinates}
    \end{equation} 
    where the operator $\nint*{\cdot}$ represents an element-wise rounding operation and the operator $\oslash$ represents the element-wise Hadamard division operator.

    When describing geometric features in the discrete 4D light field, it is often useful to reference fractional samples not belonging to its discrete grid. To this end, normalised continuous coordinates $\bar{\vc{r}}\in\set{M}\times\set{K}\subset\mathbb{R}^{4}$ are defined such that:
        \begin{equation}
		\bar{\vc{r}}=\nmatrix{\bar{\vc{u}};\bar{\vc{s}}}= 
        \nmatrix{
			\vc{u}\oslash\Delta \vc{u}+\vc{m}_{\rm r};
			\vc{s}\oslash\Delta \vc{s}+\vc{k}_{\rm r}
		} .
        \label{eq:normalization-function}
    \end{equation}  
    
From Equations~\eqref{eq:theta}, \eqref{eq:discrete-lf} and \eqref{eq:normalization-function}, Equation~\eqref{eq:4dppp} can be rewritten as:
\begin{eqnarray}
		\bar{\vc{u}} &=& \vc{\eta}\odot\left(\bar{\vc{s}} - \vc{k}_{\rm r}\right)\tan\theta + D \frac{\vc{\xi}^{(x\times y)}(\vc{x})}{z} \oslash\Delta \vc{u} + \vc{m}_{\rm r}  \nonumber \\
        &=& \vc{\eta}\odot\left(\bar{\vc{s}} - \vc{k}_{\rm r}\right) \tan\theta + \bar{\vc{u}}_0,
 \label{eq:4dppp-short}
\end{eqnarray}    
where $\vc{\eta} = \nmatrix{\frac{\Delta s}{\Delta u},\frac{\Delta t}{\Delta v}}\T\in\mathbb{R}^{2}$ is referred to as the sampling slope distortion, and $\bar{\vc{u}}_0$ represents the pixel position where the 4D-PPP intersects the reference view. { In this Equation, the 4D-PPP is parametrized by $\bar{\vc{u}}_0$, $\bar{\vc{k}}_{\rm r}$, $\theta$ and $\vc{\eta}$. Therefore, for a discrete light field with known $\vc{\eta}$ and $\vc{k}_{\rm r}$, the 4D-PPP is uniquely specified by its orientation $\theta$ and its intersection $\bar{\vc{u}}_0$ with the reference view. In this work, one will use the shorthand notation $\mathcal{P}_{\bar{\vc{u}}_0}^\theta$ to represent a 4D-PPP in a given light field.  


Estimating the depth map of a 3D scene based on an acquired light field is thus equivalent to finding the angles $\theta$ of the 4D-PPPs corresponding to all the reference view intersections $\bar{\vc{u}}_{0}$ with integer-valued coordinates, that is, finding $\theta$ such that $\mathcal{P}_{\vc{m}_0}^\theta$, $\vc{m}_0 \in \mathbb{Z}^{2}$, is a 4D-PPP.} It is common in the literature to express the orientation of the 4D-PPP in terms of the disparity, that is, the variation {in pixel position $\bar{\vc{u}}$} relative to a unit variation in view position $\Delta\bar{\vc{s}} = \nmatrix{1,1}\T$. { Thus, from Equation~\eqref{eq:4dppp-short}, the disparity associated with $\mathcal{P}_{\bar{\vc{u}_0}}^\theta$ is given by}
\begin{equation}
    \vc{d} = \vc{\eta}\tan\theta. \label{eq:disparity}
\end{equation}

\vspace*{-0.2cm}

\subsection{The 4D Point-Projection Image} \label{sec:4dppi}

{

Each 4D-PPP $\set{P}^\theta_{\bar{\vc{u}}_0}$ represents a 3D-space point. As such, for 3D-space points belonging to a Lambertian surface, in the absence of occlusions,  as per Equation~\eqref{eq:continuous-lf}, all positions $\vc{r}$ corresponding to this 3D-space point will have the same colour $\vc{c}$. Therefore, in this case, all light field samples belonging to  4D-PPP $\set{P}^\theta_{\bar{\vc{u}}_0}$ will have the same colour $\vc{c}$. This is commonly designated as photometric consistency~\cite{Zhou2023}.

It is important to note that, for a discrete light field, there is no guarantee that a given 4D-PPP intersects the discrete grid of the light field. This is similar to the case of computing intersections of a 2D continuous line with a 2D discrete image. As such, one must allow for the interpolation of the discrete light field when computing such an intersection. In this work, separable bi-linear interpolation~\cite{Kirkland2010} is used for this end. 

As there is an infinite number of intersections between the 4D-PPP and the interpolated discrete light field, for practical applications, one must choose a finite set of samples from the interpolated light field. An approach is to sample the light field for each view $\vc{k}\in\mathbb{Z}^{2}$ in spatial position $\bar{\vc{u}}$, where $\bar{\vc{u}}$ is obtained directly from Equation~\eqref{eq:4dppp-short} with $\bar{\vc{s}} = \vc{k}$. 

}

{ The result of the above sampling of a 4D-PPP $\set{P}^\theta_{\bar{\vc{u}}_0}$ at each view $\vc{k}$  is the 4D Point-Projection Image (4D-PPI) $\vc{I}^\theta_{\bar{\vc{u}}_0}(\vc{k})$, which, from Equations~\eqref{eq:discrete-lf} and \eqref{eq:4dppp-short}}, can be defined as:
\begin{equation}
    \vc{I}^\theta_{\bar{\vc{u}}_0}(\vc{k}) =\overline{\lf}(\vc{\eta}\odot \left(\vc{k} - \vc{k}_{\rm r}\right)\tan\theta + \bar{\vc{u}}_0,\vc{k}),\\ {\rm ~for~} \vc{k}\in \set{K}\cap\set{Z}^2,
    \label{eq:ppi}
\end{equation} 
where $\set{K}\cap\set{Z}^2$ is the set of all views of the discrete 4D light field, and $\overline{\lf}$ represents the view-interpolated discrete light field.

{\subsection{Estimating the 4D-PPP Orientation}
\label{subsec:orientation-estimation}
 
 Finding the correct parameters of the 4D-PPP $\set{P}^\theta_{\bar{\vc{u}}_0}$ is equivalent to finding the depth of the corresponding 3D-space point. As such, the goal of any light field depth estimation method is to find the correct orientation of 4D-PPPs. One way to do so is to analyse the corresponding 4D-PPIs (Equation~\eqref{eq:ppi}).
 
If a 4D-PPI is not photometrically consistent, then at least one of the following is correct:
 \begin{enumerate}[(i)]
   \item the Lambertian assumption does not hold;
   \item the object is occluded in some views of the light field;
   \item the parametrisation of the 4D-PPP does not match the true position of the 3D-space point represented by the pixel at position $\bar{\vc{u}}_0$ of the reference view. 
 \end{enumerate}
 
 As such, if one assumes that a scene is Lambertian and contains no occlusions, one can estimate the 4D-PPP orientation $\theta$ corresponding to a given pixel position $\vc{m}$ in the reference view by finding $\theta$ that minimises a cost function measuring the photometric consistency of the resulting 4D-PPIs. 
 }
 
 A robust example of such a cost function is the pixel deviation~\cite{Kang2021}:
\begin{equation}
    \overline{\vc{J}}_{\rm PD}=  \frac{1}{K L}\sum_{\vc{k}\in \set{K}\cap\set{Z}^2}\abs{\vc{I}^{\theta}_{\bar{\vc{u}}_0}(\vc{k}) - \vc{I}^{\theta}_{\bar{\vc{u}}_0}(\vc{k}_{\rm r})},
    \label{eq:pixel-deviation-vec}
\end{equation}
where $K$ and $L$ are the number of discrete views of the light field along the horizontal and vertical directions, respectively, $\set{K}\cap\set{Z}^2$ is the set of all views of the discrete 4D light field, and $\abs{.}$ is the element-wise norm operator, such that for $\vc{v} = \nmatrix{v_0,\dots,v_i}\T$, $\abs{\vc{v}} = \nmatrix{\abs{v_0},\dots,\abs{v_i}}\T$. Considering $\overline{\vc{J}}_{\rm PD} = \nmatrix{J^{\rm R}_{\rm PD},J^{\rm G}_{\rm PD}, J^{\rm B}_{\rm PD}}\T$, the scalar cost is defined as the average of the  cost of the three colour channels, that is:
\begin{equation}
    J_{\rm PD} = \frac{1}{3}\left(J^{\rm R}_{\rm PD} + J^{\rm G}_{\rm PD} + J^{\rm B}_{\rm PD}\right).
    \label{eq:pixel-deviation}
\end{equation}

Note that if the 4D-PPI is photometrically consistent, then its samples  $\vc{I}^{\theta}_{\bar{\vc{u}}_0}(\vc{k})$ are equal for all view indexes $\vc{k}\in\set{K}\cap\set{Z}^2$, and thus $\overline{\vc{J}}_{\rm PD}=0$.

\vspace*{-0.2cm}

\subsection{Known Limitations of 4D-PPP-based Depth Estimation}
\label{subsec:known-limitations}

Approaches based on photometric consistency, such as the one employing the 4D-PPP and the cost described by Equation~\eqref{eq:pixel-deviation-vec}, are accurate for a large percentage of situations. However, such approaches have known limitations that require depth estimation algorithms to base their results explicitly or implicitly on different heuristics and more complex models. {The most important of such limitations are related to occlusions, low variance in the image texture, inconsistencies in surface reconstruction and non-Lambertian scenes. The remainder of this section goes further in depth into these limitations apart from the ones related to non-Lambertian scenes, which fall outside the scope of this article.
}

\subsubsection{Low Variance in the Imaged Texture} \label{subsec:low_variance}

From the discussion in Subsection~\ref{subsec:orientation-estimation} above, 4D-PPIs with the correct orientation will have near-constant colour outside occlusion and non-Lambertian situations. { 

However, the converse is not true, since 
in regions of a scene with low colour variance, even 4D-PPIs with incorrect orientations can reveal photometric consistency. 
To demonstrate this fact, let us assume a 4D-PPI obtained from a hypothetical 4D-PPP $\set{P}^{\theta}_{\vc{m}_0}$, with an incorrect angle $\theta$ such that $\tan\theta=\tan\theta_{\rm c}+\delta$. 
From Equation~\eqref{eq:ppi}:
\begin{eqnarray}
        \vc{I}^\theta_{\vc{m}_0}(\vc{k}) 
        &=& \overline{\lf}(\vc{\eta}\odot \left(\vc{k} - \vc{k}_{\rm r}\right)\tan\theta_{\rm c} +\vc{\eta}\odot \left(\vc{k} - \vc{k}_{\rm r}\right)\delta+\vc{m}_0,\vc{k})  \nonumber \\
        &=&  \vc{I}^{\theta_{\rm c}}_{\vc{m}_0+\vc{\eta}\odot \left(\vc{k} - \vc{k}_{\rm r}\right)\delta}(\vc{k}) = \vc{I}^{\theta_{\rm c}}_{\vc{m}_0+ \boldsymbol{\epsilon}}(\vc{k}), 
        \label{eq:ppi_wrong_theta}
 \end{eqnarray}
 
    The 4D-PPI $\vc{I}^\theta_{\vc{m}_0}(\vc{k})=\vc{I}^{\theta_{\rm c}}_{\vc{m}_0+\boldsymbol{\epsilon}}(\vc{k})$ in Equation~\eqref{eq:ppi_wrong_theta} can be interpreted as 
    an image that, for each view, samples a 4D-PPP with the correct orientation $\theta_{\rm c}$, but intersects the reference view at a wrong position $\vc{m}_0+\boldsymbol{\epsilon}$.  Furthermore, from Equation~\eqref{eq:ppi_wrong_theta} the position error $\boldsymbol{\epsilon}$ is given by $\vc{\eta}\odot \left(\vc{k} - \vc{k}_{\rm r}\right)\delta$ and thus increases with the orientation error $\delta$.

    Therefore, if the reference view has a large variance in the neighbourhood of $\vc{m}_0$, $\vc{I}^\theta_{\vc{m}_0}(\vc{k})=\vc{I}^{\theta_{\rm c}}_{\vc{m}_0+\boldsymbol{\epsilon}}(\vc{k})$ will not demonstrate photometric consistency and a large cost will be attributed to the wrong $\theta$. However, if the variance in the neighbourhood is small, $\vc{I}^\theta_{\vc{m}_0}(\vc{k})=\vc{I}^{\theta_{\rm c}}_{\vc{m}_0+\boldsymbol{\epsilon}}(\vc{k})$ may demonstrate photometric constancy even for larger errors $\delta$.}  
    

\subsubsection{Occlusions} \label{subsec:occlusions}  
Not all real-world points are visible in all views of the light field. { For example, a 3D scene point $\vc{x}_0$ can lie behind a second point $\vc{x}_{\rm occ}$ such that this point occludes it from the camera for a given view.

In a 4D light field, $\vc{x}_0$ and $\vc{x}_{\rm occ}$ are represented by 4D-PPPs $\set{P}^{\theta_0}_{\vc{m}_0}$  and $\set{P}^{\theta_{\rm occ}}_{\bar{\vc{u}}}$, respectively. An occlusion is represented by the \textit{intersection} of these two 4D-PPPs. From Equation~\eqref{eq:4dppp-short}, such an intersection is described by the following equation:
	\begin{equation}
			\vc{\eta}\odot (\bar{\vc{s}}-\vc{k}_{\rm r})\tan\theta_0+\vc{m}_0 = \vc{\eta}\odot(\bar{\vc{s}}-\vc{k}_{\rm r})\tan\theta_{\rm occ}+\bar{\vc{u}}_0.
		  \label{eq:occlusion}
	\end{equation}
where $\bar{\vc{s}}$ is the view where the 4D-PPPs intersect and thus the view where point $\vc{x}_0$ is occluded by $\vc{x}_{\rm occ}$.


It is important to note that since} $\vc{x}_{\rm occ}$ occludes $\vc{x}_{0}$ then the depth $z_{\rm occ}$ of $\vc{x}_{\rm occ}$ is necessarily smaller than the depth $z_{0}$ of $\vc{x}_{0}$. From Equation~\eqref{eq:theta} this implies that
\begin{equation}
   \begin{cases}
       \theta_{\rm occ} < \theta_0 \quad {\rm if~} D>0, \\
       \theta_{\rm occ} > \theta_0 \quad {\rm if~} D<0. \label{eq:cases_theta_occ}
   \end{cases}
\end{equation}
In this text, without loss of generality, it can be assumed that $D<0$, which means that the camera centres are located between the sensor and the object.

\begin{figure}[t]
	\centering
	\includegraphics[width = 0.25\textwidth]{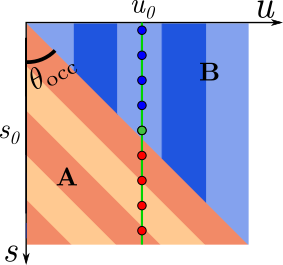}
   \caption{A diagram of an $s\times u$ EPI segment showcasing two different regions of constant depth. Region A has an orientation $\theta_{\rm occ}$ and occludes region B, which has orientation $\theta_0 = 0$. The green line represents the intersection of the EPI with the 4D-PPP crossing the central view at the point $(u_0,s_0)$ (in green). Blue circles represent unoccluded samples of the 4D-PPP, and red circles its occluded samples.} 
   \label{fig:occlusion-diagram}
\end{figure}

Figure \ref{fig:occlusion-diagram} illustrates this situation in an $s\times u$ EPI that depicts a hypothetical situation with two regions of constant depth and some texture.
Region A represents an occluding region with a constant 4D-PPP angle equal to $\theta_{\rm occ}$ while region B is a partially occluded region with a 4D-PPP angle $\theta_0 = 0$.

The green point represents a light field position $\vc{p}$ for which the orientation $\theta$ is estimated. If one supposes that $\theta(\vc{p}) =  \theta_0 = 0$, that is the correct angle for Region B, the 4D-PPP (whose intersection with the EPI is represented as a green line) will intersect both Regions A and B. The samples of views with $s \geq s_0$ (red dots) will correspond to the occluding region, and those for $s<s_0$ (blue dots) will correspond to the occluded region. This implies that the 4D-PPI corresponding to the correct 4D-PPP orientation does not present photometric consistency. Therefore, any method solely based on photometric consistency may lead to inaccurate results near occluded regions.

\subsubsection{Inconsistencies in Surface Reconstruction}
\label{subsec:surface_inconsistencies}

{ Due to image noise, calibration errors and the issues described in Subsubsection \ref{subsec:low_variance}, it is very likely that there will be a non-zero orientation estimation error for each sample. If the orientation estimations for neighbouring samples are uncorrelated, small estimation errors can result in much larger errors in the geometry of the corresponding 3D scenes, generating inconsistencies such as} smooth surfaces appearing rugged or stair-case effects in slanted planes. These inconsistencies are undesirable when estimating surface normals from 3D reconstructions obtained from light fields. In order to assess these, { the \textit{4D Lightfield Benchmark} \cite{Johansen2017} proposes the median angle error (MAE) metric to measure surface normal accuracy. This measure consists of the median of the angle differences, in degrees, between the surface normals estimated from a given depth map and the provided ground truth in a neighbourhood. 

Algorithms for calculating surface normals require a compromise: using the fewest points possible results in well-localised normals, but such normals are very susceptible to any error in the points used. Alternatively, using a larger set of points leads to increased robustness to error but tends to average out geometrical details, thus generating ill-defined borders. 

The algorithm for surface normal computation using the least number of points} requires accurate estimations of the 3D coordinates of three 3D-space points: one corresponding to pixel coordinates $\vc{m}$ at a given view, $\vc{x} (\vc{m}) = \nmatrix{x(\vc{m}),y(\vc{m}),z(\vc{m})}\T$, and two of its neighbours, corresponding to coordinates  $\vc{x}(\vc{m}+\vc{e}^{\rm h})$ and $\vc{x}(\vc{m}+\vc{e}^{\rm v})$, where
\begin{equation}
    \vc{e}^{\rm h} = \nmatrix{1,0}\T \quad {\rm and}, \quad 
    \vc{e}^{\rm v} = \nmatrix{0,1}\T \label{eq:e_hv}.
\end{equation}

The orientation of the normal of the surface containing $\vc{x}(\vc{m})$, $\vc{x}(\vc{m}+\vc{e}^{\rm h})$ and $\vc{x}(\vc{m}+\vc{e}^{\rm v})$ is
$\vc{\nu}(\vc{m}) = \frac{\vc{\tau}^{\rm n}(\vc{m})}{\|\vc{\tau}^{\rm n}(\vc{m})\|}$, where the non-normalized surface normal vector $\vc{\tau}^{\rm n}(\vc{m})$ is obtained through the cross product:
\begin{equation}
    \vc{\tau}^{\rm n}(\vc{m}) = \vc{\tau}^{\rm h}(\vc{m}) \times \vc{\tau}^{\rm v}(\vc{m}),
    \label{eq:normal-map}
\end{equation}
where 
\begin{equation}
\begin{aligned}
    \vc{\tau}^{\rm h}(\vc{m}) &= \vc{x}(\vc{m}+\vc{e}^{\rm h})-\vc{x}(\vc{m}),  \\
    \vc{\tau}^{\rm v}(\vc{m}) &= \vc{x}(\vc{m}+\vc{e}^{\rm v}) - \vc{x}(\vc{m}). 
\end{aligned}
\label{eq:tau_simple}
\end{equation}

{ A more robust approach is the use of difference kernel filters~\cite{Nakagawa2015}, which smooth the 3D space} points while computing $\vc{\tau}^{\rm h}$ and $\vc{\tau}^{\rm v}$, yielding:
\begin{equation}
    \begin{aligned}
        \vc{\tau}^{\rm h}(\vc{m}) &= \sum_{\vc{i}\in {\cal W}} \vc{x}(\vc{m}+\vc{i}) g^{\rm h}(\vc{i}),\\
        \vc{\tau}^{\rm v}(\vc{m}) &= \sum_{\vc{i}\in {\cal W}} \vc{x}(\vc{m}+\vc{i}) g^{\rm v}(\vc{i}),\\
    \end{aligned}
    \label{eq:tangent-map}
\end{equation}
where $\vc{i}\in{\cal W}\subset \mathbb{Z}^2$ is the index of a difference kernel filter and $g^{\rm h}(\vc{i})$ and $g^{\rm v}(\vc{i})$ are the coefficients of difference kernel filters along the horizontal and vertical directions inside a view, respectively. Usually, $g^{\rm h}(\vc{i})$ and $g^{\rm v}(\vc{i})$ have some symmetry in the sense that $g^{\rm v}([i~j]) = g^{\rm h}([j~i])$, $\forall[i~j]\in\mathbb{Z}^{2}$. { This is the approach used to obtain the surface normal maps used for the calculation of the MAE metric proposed in \cite{Johansen2017}, with $g^{\rm h}(\vc{i})$ and $g^{\rm v}(\vc{i})$ being $3\times 3$ Scharr filters.} 

\section{The Proposed Cost Model}
\label{sec:cost-model}
{In this section, we propose a robust cost model capable of addressing directly and indirectly the known limitations of 4D-PPP-based light field depth estimation mentioned in Subsection \ref{subsec:known-limitations}.

Our goal is to obtain a scalar cost for any given 4D-PPP $\set{P}^{\theta}_{\vc{m}_0}$  that indicates how well this plane fits the model. As the main interest is calculating the orientation for each pixel of reference view $\vc{k}_r$, this can be established as a function $J(\theta;\vc{m}_0,\theta_{\rm map}(\vc{m}))$ where:
\begin{itemize}
    \item $\theta$ is the candidate orientation.
    \item $\vc{m}_0$ is a pixel location in the reference view $\vc{k}_{\rm r}$.
    \item $\theta_{\rm map}(\vc{m})$ is a map that gives the current best estimate of the orientation at each location of the pixel in the reference view $\vc{k}_{\rm r}$.
\end{itemize}
This cost function can be further divided into a weighted sum of three terms as:}
\begin{equation}
    J(\theta;\vc{m}_0,\theta_{\rm map}(\vc{m})) = J_{\rm oa}(\theta;\vc{m}_0,\theta_{\rm map}(\vc{m})) + \lambda J_{\rm coc}(\theta;\vc{m}_0,\theta_{\rm map}(\vc{m})) + \gamma J_{\rm pg}(\theta;\vc{m}_0,\theta_{\rm map}(\vc{m})),
\label{eq:cost} 
\end{equation}
where:
\begin{itemize} 
   \item $J_{\rm oa}(\theta;\vc{m}_0,\theta_{\rm map}(\vc{m}))$ is a 4D occlusion-aware data cost;
   \item $J_{\rm coc}(\theta;\vc{m}_0,\theta_{\rm map}(\vc{m}))$ is a colour-orientation congruence cost;
   \item $J_{\rm pg}(\theta;\vc{m}_0,\theta_{\rm map}(\vc{m}))$ is a planar geometry cost. 
\end{itemize}

The remainder of the section details each of these costs.

\subsection{4D Occlusion-Aware Data Cost}

As addressed in Subsection \ref{subsec:known-limitations}, occlusions constitute some of the main difficulties in obtaining accurate 4D-PPP orientation estimates from data costs, such as the variance or the pixel deviation of potential 4D-PPIs from Equation~\eqref{eq:pixel-deviation-vec}. { 

As discussed in Subsection~\ref{subsec:occlusions}, these difficulties brought by occlusions arise because some of the samples in a 4D-PPI are the result of occlusions. This work proposes to deal with occlusions by employing an occlusion-aware cost that can be computed, given a 4D-PPP $\set{P}^{\theta}_{\vc{m}_0}$, using the following three-step process:
\begin{enumerate}[1.]
    \item Compute the 4D-PPI $\vc{I}^{\theta}_{\vc{m}_0}(\vc{k})$ as described in Subsection~\ref{sec:4dppi}.
    \item Estimate the set $\overline{\set{O}}$ of sample positions $\vc{k}$ of the 4D-PPI that do not result from occlusions (Algorithm \ref{alg:data_cost}).
    \item Calculate a cost metric only using the samples in $\overline{\set{O}}$ as:
    \begin{equation}
    J_{\rm oa}(\theta,\vc{m}_0) = \frac{1}{3}\left(J_{\rm oa}^{\rm R}+J_{\rm oa}^{\rm G}+J_{\rm oa}^{\rm B}\right),
    \label{eq:data-cost}
\end{equation}
where $\overline{\vc{J}}_{\rm oa}(\theta,\vc{m}_0) = \nmatrix{J_{\rm oa}^{\rm R},J_{\rm oa}^{\rm G},J_{\rm oa}^{\rm B}}\T$ is given by:
    \begin{equation}
     \overline{\vc{J}}_{\rm oa}(\theta,\vc{m}_0) =  \dfrac{1}{|\overline{\set{O}}|}\displaystyle\sum_{\vc{k}\in \overline{\set{O}}}\abs{\left(\vc{I}^{\theta}_{\vc{m}_0}(\vc{k}) - \vc{I}^{\theta}_{\vc{m}_0}(\vc{k}_{\rm r}) \right)},
\end{equation}
where $\vc{k}_{\rm r}$ is the index of the reference view.

\end{enumerate}


The major hurdle in this process is the second step: accurately detecting which samples of the 4D-PPI are members of $\overline{\set{O}}$. 
 This step is detailed in Algorithm \ref{alg:data_cost} below, that first determines the orientations $\theta_{\rm occ}$ of all 4D-PPPs that could potentially occlude $\set{P}^{\theta}_{\vc{m}_0}$. Each of these orientations is then tested for each sample position $\vc{k}\in\set{K}\cap\set{Z}^2$ of the 4D-PPI $\vc{I}^{\theta}_{\vc{m}_0}(\vc{k})$. 

}

{
\begin{algorithm} 
\caption{\textbf{Determination of Unoccluded 4D-PPI Samples}} \label{alg:data_cost}
\vspace*{-1.8em}
\end{algorithm}
\vspace*{0.4em}
\begin{enumerate}[I.]
\item \textit{Inputs:}
\begin{itemize}
    \item A discrete orientation map of the reference view $\theta_{\rm map}(\vc{m})$.
    \item The maximum orientation of the light field $\theta_{\rm max}$, associated with the minimum depth of the scene.
    \item The current reference view position $\vc{m}_0$.
    \item The candidate orientation $\theta$ for the current 4D-PPP.
\end{itemize}
\item \textit{Outputs:}
\begin{itemize}
    \item The set $\overline{\set{O}}$, containing all view indexes $\vc{k}$ corresponding to unoccluded colour samples of the 4D-PPI.
\end{itemize}
\item \textit{Determination of indexes $\vc{k}$ of the unoccluded samples of the 4D-PPI}
\begin{enumerate}[i.]

\item  Initialize $\overline{\set{O}} = \set{K}\cap\set{Z}^2$ (Equation \eqref{eq:lf-boundaries}).
\item Compute $\bar{\theta}(\bar{\vc{u}})$, the interpolated depth map of the reference view, by applying bilinear interpolation to the discrete orientation map $\theta_{\rm map}(\vc{m}_0)$.
\item Equal $\set{T}$ to the set of orientations $\theta_{\rm occ}$ for which Equation \eqref{eq:occlusion} has a solution for $\bar{\vc{u}}_0 \in \set{M}$ and $\bar{\vc{s}} \in \set{K}$, and
\begin{equation}
    \theta_{\rm occ} \in (\theta,\theta_{\rm max}]\cap \{\theta \mid \theta_{\rm occ} = \theta_{\rm map}(\vc{m}),\ \forall \vc{m} \in \set{M} \},
\end{equation}
where $\set{M}$ and $\set{K}$ are defined in Equation \eqref{eq:lf-boundaries}.

   

    \item For all view indexes $\vc{k}\in\set{K}\cap\set{Z}^2$ and for all ${\theta_{\rm occ}\in\set{T}}$, do
    \begin{enumerate}[a.]
        \item  Compute $\bar{\vc{u}}_0 = \bar{\vc{u}}_{\rm occ}$ by solving  Equation~\eqref{eq:occlusion} for $\bar{\vc{s}} = \vc{k}$.
        \item   Compute $\bar{\vc{s}} = \bar{\vc{s}}_{\rm occ} $ by solving  Equation~\eqref{eq:occlusion} for $\bar{\vc{u}}_0 = \bar{\vc{u}}_{\rm occ}$ and $\theta_{\rm occ} = \bar{\theta}(\bar{\vc{u}}_{\rm occ})$.
        
        \item      If ${\|\vc{k} -  \bar{\vc{s}}_{\rm occ}}\|_\infty < \frac{1}{2}$ exclude $\vc{k}$ from $\overline{\set{O}}$.
    
        \end{enumerate}

\end{enumerate}
\hrule\vspace{0.3em}
\end{enumerate}
\vspace{0.2em}

}

 \subsection{Colour-Orientation Congruence Cost} \label{subsubsubsec:cocc}
 
 { An occlusion in a light field corresponds to an edge in its depth map, which in almost all cases corresponds to an edge in the corresponding views. Therefore, a correctly estimated depth map of a light field view tends to have the orientations of its edges congruent to the orientations of the edges of its corresponding view.  Considering the above and the fact that according to Eq.~\eqref{eq:theta} the orientation $\theta$ of a 4D-PPP is equivalent to the depth of its corresponding 3D-space point, this paper proposes to add a Colour-Orientation Congruence term ($J_{\rm coc}$)  to the data cost, that measures the degree of agreement between the edge map of the reference view and the 4D-PPP orientation map. 
 
The Colour-Orientation Congruence term $J_{\rm coc}$ is based on the smoothness cost from Schilling \textit{et al.}~\cite{Schilling2018}.
It is computed by comparing a candidate orientation value $\theta$ with a smoothed orientation $\theta_s(\vc{m}_0)$, which is obtained via a guided weighted filter that keeps edges in the orientation map congruent with edges in the reference view of the light field. Given a candidate orientation $\theta$, $J_{\rm coc}$ is computed for position $\vc{m}_0$ of the reference view $\vc{k}_{\rm r}$  as:
\begin{equation}
    J_{\rm coc}(\theta,\vc{m}_0,\theta_{\rm map}(\vc{m})) = \left(\tan\theta - \tan(\theta_{\rm s}(\vc{m}_0)\right)^2,
    \label{eq:smoothing-term}
\end{equation}
where $\theta_{\rm s}(\vc{m})$ is a smoothed version of $\theta_{\rm map}(\vc{m})$ that tries to preserve edge information in the orientation map by maintaining congruency with the color (edge) information at the centre view of the light field. It is obtained as:
\begin{equation}
\theta_{\rm s} (\vc{m}_0) = \tan^{-1}\left(\dfrac{\displaystyle\sum_{\vc{m}\in {\cal W^{\rm c}}} \chi(\vc{m}) \theta_{\rm map}(\vc{m})}{\displaystyle\sum_{\vc{m}\in {\cal W}^{\rm c}} \chi(\vc{m})}\right),
    \label{eq:smoothing-filter}
\end{equation} 
where ${\cal W^{\rm c}}$ is a window around position $\vc{m}_0$ in the reference view $\vc{k}_{\rm r}$ and $\chi(\vc{m})$ is the weight of the orientation sample at $\vc{m}$.  The weight $\chi(\vc{m})$ is computed as:
\begin{equation}
\chi(\vc{m}) = 
\begin{cases}
          \mathrm{max}\left\{\epsilon_\theta,\sqrt{\Delta_\theta(\vc{m})^2+\Delta_c(\vc{m}) \Delta_\theta(\vc{m})}\right\}^{-1},& \Delta_c(\vc{m}) \leq \tau_{\rm c} {\rm ~and~} \Delta_\theta(\vc{m}) \leq \tau_{\rm o}, \\
          \mathrm{max}\left\{\epsilon_\theta,\sqrt{\Delta_c(\vc{m})^2 +\Delta_\theta(\vc{m})^2}\right\}^{-1}, & \Delta_c(\vc{m}) \leq \tau_{\rm c} {\rm ~and~} \Delta_\theta(\vc{m}) > \tau_{\rm o}, \\
          0, & {\rm elsewhere},
\end{cases}
\end{equation}
where $\epsilon_\theta$ and $\tau_{\rm c}$ are determined empirically, $\tau_{\rm o}$ is the dynamic range of $\tan\theta$ for the light field being processed, and $\Delta_{\theta}(\vc{m})$ and $\Delta_{c}(\vc{m})$ are the orientation and colour differences, respectively. They are defined as:
\begin{eqnarray} 
    \Delta_{\theta}(\vc{m}) &=& \rho_{\theta}|\tan \theta_{\rm map}(\vc{m}) - \tan\theta|,\\
    \Delta_{c}(\vc{m}) &=& \rho_{c}\|\lf(\vc{m}_{0},\vc{k}_{\rm r}) - \lf(\vc{m},\vc{k}_{\rm r})\|,
\end{eqnarray} 
where $\lf(\vc{m},\vc{k})$ is the discrete light field as defined in Equation \eqref{eq:discrete-lf}, and the weights $\rho_{\theta}$ and $\rho_{c}$ are defined empirically.

A small $J_{\rm coc}$, therefore, implies that the candidate $\theta$ is congruent with the light field colour variations. In contrast, a high cost implies that the candidate $\theta$ would lead to abrupt transitions in the orientation map that are not matched by the expected edges in the light field.}

 \subsection{Planar Geometry Cost}
 \label{subsec:planar-geometry-cost}
 
{ As discussed in Section \ref{subsec:known-limitations}, 
proper 3D reconstruction requires the inconsistencies in surface reconstruction to be addressed when building the data cost. However, such inconsistencies are not taken into consideration by the cost terms $J_{\rm oa}$ (Equation~\eqref{eq:data-cost}) and $J_{\rm coc}$ (Equation~\eqref{eq:smoothing-term}). To address this issue, this paper proposes the introduction of a novel planar geometry cost term  $J_{\rm pg}(\theta,\vc{m}_0, \theta_{\rm map}(\vc{m}))$ to the cost model. It tests the impact on the accuracy of surface normals of a candidate orientation $\theta$ associated to sample $\vc{m}_{0}$ of the reference view given the current orientation map $\theta_{\rm map}(\vc{m})$, $\vc{m}\in {\cal M}\cap\mathbb{Z}^{2}$. 

The surface normals can be estimated from the 3D space-point map $\vc{x}_{\rm map}(\vc{m})$, $\vc{m}\in {\cal M}\cap\mathbb{Z}^{2}$, using Equations~\eqref{eq:normal-map} to \eqref{eq:tangent-map}; the 3D space-point map $\vc{x}_{\rm map}(\vc{m})$ can be derived from the light field samples and the orientation map $\theta_{\rm map}(\vc{m})$ using Equations~\eqref{eq:4dppp} to \eqref{eq:4dppp-short}. Given a candidate orientation $\theta$ associated to the sample $\vc{m}_{0}$ of the reference view, one computes the corresponding candidate 3D-space point $\vc{x}_{\rm cnd}(\vc{m}_{0})$, that, together with the current 3D-space point map $\vc{x}_{\rm map}(\vc{m})$ for $\vc{m}$ in a neighbourhood of $\vc{m}_{0}$, is used to compute the surface normals $\vc{\nu}(\vc{m})$ associated with the candidate orientation $\theta$.  
This is done using Equations~\eqref{eq:normal-map} and \eqref{eq:tangent-map} with $\vc{x}(\vc{m}_{0})=\vc{x}_{\rm cnd}(\vc{m}_{0})$ and $\vc{x}(\vc{m})=\vc{x}_{\rm map}(\vc{m})$ for $\vc{m}\neq \vc{m}_{0}$.  


The planar geometry cost $J_{\rm pg}$ is computed based on the fact that a wrongly estimated candidate orientation $\theta$ at $\vc{m}_{0}$ will generate a wrongly estimated $\vc{x}_{\rm cnd}(\vc{m}_{0})$, which will introduce errors in the estimation of the surface normals $\vc{\nu}(\vc{m})$ in the neighbourhood of $\vc{m}_{0}$. In planar surfaces, these errors will depend on the size ${\cal W}$ of the kernels $g^h$ and $g^v$ in Equation~\eqref{eq:tangent-map}. Smaller kernels will lead to normal estimates that are more sensitive to errors in $\vc{x}_{\rm cnd}(\vc{m}_{0})$ than larger kernels. One then computes two normal estimates: $\vc{\nu}_{\rm sm}(\vc{m}_{0})$ using small kernels $g^{\rm h}_{\rm sm}(\vc{i})$ and $g^{\rm v}_{\rm sm}(\vc{i})$, and a more robust $\vc{\nu}_{\rm lg}(\vc{m}_{0})$ using larger kernels $g^{\rm h}_{\rm lg}(\vc{i})$ and $g^{\rm v}_{\rm lg}(\vc{i})$. In a 3D neighbourhood that is approximately planar, if the 3D space-point map is accurate, then $\vc{\nu}_{\rm sm}(\vc{m}_{0})$ and $\vc{\nu}_{\rm lg}(\vc{m}_{0})$ tend to have the same orientations. As such, a candidate orientation $\theta$ that minimises the error in the surface normal estimation tends to be one that minimises the angle between $\vc{\nu}_{\rm sm}(\vc{m}_{0})$ and $\vc{\nu}_{\rm lg}(\vc{m}_{0})$, and the planar geometry cost $J_{\rm pg}$ is computed as this angle. 



In this work, the small kernels are simple difference kernels such that \linebreak $g^{\rm v}_{\rm sm}([i~j]) = g^{\rm h}_{\rm sm}([j~i])$, $\forall[i~j]\in\mathbb{Z}^{2}$, and
\begin{equation}
    g^{\rm h}_{\rm sm}(\vc{i})= 
    \begin{cases}
        i, & \vc{i} = [i,0] {\rm~and~} \|\vc{i}\|=1, \\
        0, & {\rm ~otherwise}.
    \end{cases} \label{eq:small_kernel}
\end{equation}
The large kernels are Gaussian difference kernels such that $g^{\rm v}_{\rm lg}([i~j]) = g^{\rm h}_{\rm lg}([j~i])$, $\forall[i~j]\in\mathbb{Z}^{2}$, and
\begin{equation}
    g^{\rm h}_{\rm lg}(\vc{i}) = ie^{-\frac{\|\vc{i}\|^2}{(2\delta_{a}+1)^{2}}}, \quad  \|\vc{i}\|_{\infty} \leq \delta_{a},
\end{equation}
where $\vc{i} = [i,j]$,  $\|\vc{i}\|_{\infty} = \max\{|i|,|j|\}$ and  $\delta_{a}\in\mathbb{N}$ is a parameter of the algorithm. 

%

However, there may be situations when the neighbourhood of the 3D space-point $\vc{x}_{\rm cam}(\vc{m}_{0})$ is not planar. In these cases, the use of kernels $g^{\rm v}_{\rm lg}(\vc{i})$ and $g^{\rm h}_{\rm lg}(\vc{i})$ with large values of the parameter $\delta_{a}$ may lead to inconsistent results. To address these cases while maintaining the good properties of the normal estimate $\vc{\nu}_{\rm lg}(\vc{m}_0)$ for planar regions, one can derive a robust estimate $\vc{\nu}_{\rm rob}(\vc{m}_0)$ as the average of all the normals $\vc{\nu}_{\rm lg}(\vc{m})$ whose orientation is sufficiently close to the one of $\vc{\nu}_{\rm lg}(\vc{m}_0)$. If $\beta(\vc{m}_{i},\vc{m}_{j})=\cos^{-1}\langle\vc{\nu}_{\rm lg}(\vc{m}_{i}),\vc{\nu}_{\rm lg}(\vc{m}_{j})\rangle$ (where $\langle\cdot,\cdot\rangle$ is the inner product)  is the angle between the normals associated to samples $\vc{m}_{i}$ and $\vc{m}_{j}$ of the reference view, then $\vc{\nu}_{\rm rob}(\vc{m}_{0})$ can be computed as:
%
\begin{equation}
    \vc{\nu}_{\rm{rob}}(\vc{m}_{0}) = \kappa\sum_{\substack{\beta(\vc{m},\vc{m}_{0}) < \tau_{a}\mu \\ \vc{m}\in{\cal W}^{\rm avg} }}\vc{\nu}_{\rm lg}(\vc{m}), \label{eq:nu_rob}
\end{equation}
where $\kappa$ is set so that $\vc{\nu}_{\rm{rob}}(\vc{m}_{0})$ has unit norm,  ${\cal W}^{\rm avg}$ is a window around $\vc{m}_0$, $\tau_{\rm a}$ is an empirically defined parameter, and $\mu$ is the average angle between $\vc{\nu}_{\rm lg}(\vc{m}_0)$ and the surface normals belonging to ${\cal W}^{\rm avg}$, that is, 
\begin{equation}
        \mu = \frac{1}{|{\cal W}^{\rm avg}|}\sum_{\vc{m}\in\cal{W}^{\rm avg}}  \beta(\vc{m},\vc{m}_{0}). \label{eq:mu_nu}
    \end{equation}

Despite the increased robustness of this average, it does not guarantee accuracy if ${\cal W}^{\rm avg}$ encompasses samples of the light field that correspond to a non-planar region of the 3D space. This can result in high costs even for accurate $\theta$ candidates. To that end, a new orientation $\theta_{\mu}$ is estimated from the robust normal estimate $\vc{\nu}_{\rm rob}(\vc{m}_{0})$ and all neighbouring samples $\vc{x}_{\rm map}(\vc{m})$ with $\vc{m}\in{\cal W}^{\rm avg}$. If the difference between $\theta_{\rm \mu}$ and the current orientation estimate $\theta_{\rm map}(\vc{m}_0)$ is above an empirically defined threshold the region is considered non-planar, and the cost $J_{\rm pg}$ is set to 0. The new orientation $\theta_{\mu}$ is computed by the following four step process:
\begin{enumerate}
    \item For all $\vc{m}$ such that $\beta(\vc{m},\vc{m}_{0}) < \tau_{a}\mu$ and $\vc{m}\in{\cal W}^{\rm avg}$ (see Equations~\eqref{eq:nu_rob} and \eqref{eq:mu_nu}), obtain the equation of the plane with surface normal $\vc{\nu}_{\rm rob}(\vc{m}_0)$ that passes through $\vc{x}_{\rm map}(\vc{m}$) as
    \begin{equation}
        \langle\vc{x} - \vc{x}_{\rm map}(\vc{m}),\vc{\nu}_{\rm rob}(\vc{m}_0) \rangle = 0. \label{eq:plane_with_normal}
    \end{equation}
    Note that, for different $\vc{x}_{\rm map}(\vc{m})$, these equations differ only by the offset parameter $\langle\vc{x}(\vc{m}),\vc{\nu}_{\rm rob}(\vc{m}_0)\rangle$.
    \item Compute the average $o_{\mu}$ of all the offset parameters $\langle\vc{x}(\vc{m}),\vc{\nu}_{\rm rob}(\vc{m}_0)\rangle$ of the planes obtained in step 1 above. 
    \item Considering the plane with surface normal $\vc{\nu}_{\rm rob}(\vc{m}_{0})$ and offset $o_{\mu}$,
    \begin{equation}
        \langle\vc{x},\vc{\nu}_{\rm rob}(\vc{m}_0) \rangle = o_{\mu}, \label{eq:plane_with_normal_average_offset}
    \end{equation}
    compute the 3D point $\vc{x}_\mu = \nmatrix{x_{\mu},y_{\mu},z_{\mu}}^{T}$ belonging to this plane that would be imaged at position $\vc{m}_0$ of the reference view of the light field. This is done by solving the system given by Equations~\eqref{eq:4dppp} and \eqref{eq:plane_with_normal_average_offset}.
    \item Use Equation~\eqref{eq:theta} for $z=z_{\mu}$ to compute the angle $\theta_\mu=\theta$.
\end{enumerate}

If $\theta_{\mu}$ differs too much from the sample $\theta_{\rm map}(\vc{m}_0)$ of the current 4D-PPP orientation map, then the plane geometry cost is equal to zero; otherwise, it is given by the angle difference between the robust normal estimate $\vc{\nu}_{\rm rob}(\vc{m}_{0})$ and the normal estimated using the small kernel from Equation~\eqref{eq:small_kernel}, that is: 
\begin{equation}
\overline{J}_{\rm pg}(\theta,\vc{m}_0,\theta_{\rm map}(\vc{m})) =
\begin{cases}
       \cos^{-1}\langle\vc{\nu}_{\rm rob}(\vc{m}_0),\vc{\nu}_{\rm sm}(\vc{m}_0)\rangle, &  \rm{if}\ |\tan{\theta_\mu} - \tan{ \theta_{\rm map}(\vc{m}_0)} | < \tau_\epsilon, \\
       0, & \rm{otherwise},
      
      \end{cases}
      \label{eq:pg-cost}
\end{equation} 
 where $\tau_\epsilon$ is a parameter determined empirically.
    

}

\section{Iterative Occlusion-Aware Depth Refinement}
\label{sec:ioadi}

This section presents the proposed Iterative Occlusion-Aware Depth Refinement (IOADR) algorithm { that iteratively improves on an initial 4D-PPP orientation map by minimizing the cost model presented in Section \ref{sec:cost-model}.

\begin{figure}[ht]
    \centering
    \includegraphics[width = 1.0\columnwidth]{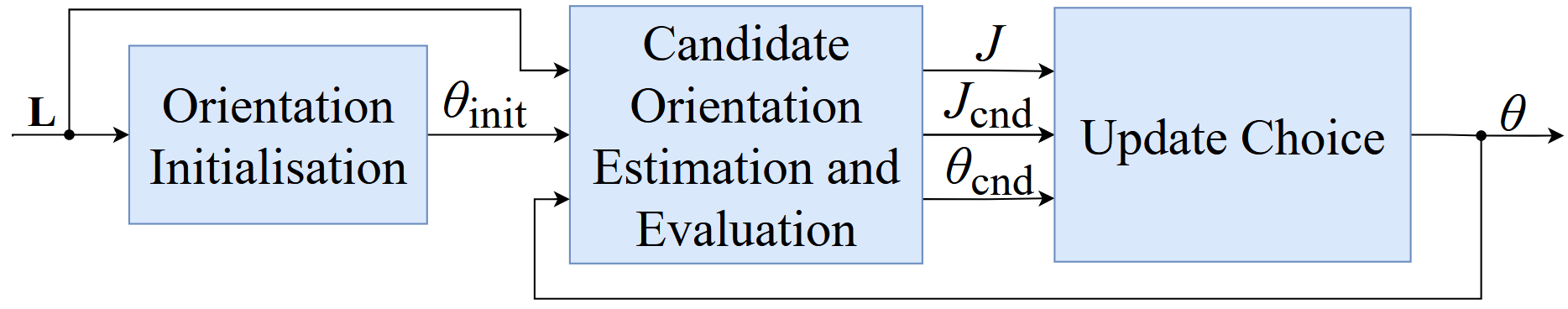}
    \caption{Diagram of the IOADR: $\lf$ represents the light field; $\theta_{\rm init}$ represents the initial orientation map; $\theta$ represents the current working 4D-PPP orientation map and $J$ its cost; $\theta_{\rm cnd}$ represents the candidate orientation and $J_{\rm cnd}$ its cost. }
    \label{fig:opt-diag}
\end{figure}

The goal of the algorithm is to iteratively compute an orientation map $\theta^{(q)}(\vc{m})$ per iteration, where $q$ is the iteration index. 
The algorithm achieves this by following an architecture that can be summarized into three major modules, as represented in Figure \ref{fig:opt-diag}:
\begin{itemize}
    \item \textit{Orientation Initialisation} --- This module provides a fast initialisation $\theta_{\rm init}(\vc{m})$ of the orientation map $(\theta^{(0)}(\vc{m})=\theta_{\rm init}(\vc{m})$) using the structure-tensor~\cite{Bigun1987} on the EPIs of the light field. More details are provided in Subsection \ref{subsec:orientation-initialization}.
    \item \textit{Candidate Orientation Estimation and Evaluation} --- For each pixel $\vc{m}_0$ of the reference view $\vc{k}_r$, this module computes the candidate orientation $\theta_{\rm cnd}$ that minimises the cost $J(\theta_{\rm cnd},\vc{m}_0,\theta^{q-1}(\vc{m})$) from Equation~\eqref{eq:cost}. The orientation $\theta_{\rm cnd}$ is chosen out of a set  $\set{C} = \{\theta^1_{\rm cnd},\cdots,\theta^N_{\rm cnd}\}$ of $N$ valid candidate orientations. More details are provided in Subsection \ref{subsec:estimate-orientation-candidates-and-costs}.
    \item \textit{Update Choice} --- This module makes a stochastic decision inspired by the simulated-annealing algorithm~\cite{bertsimas1993}: randomly, the current orientation map is either left unchanged or updated with the best candidate orientation. This increases robustness in the optimisation, preventing the algorithm from being trapped in local minima. Further details are provided in Subsection \ref{subsec:update-choice}.
\end{itemize}
}
 The IOADR algorithm continuously iterates the modules ``\textit{Candidate Orientation Estimation and Evaluation}'' and ``\textit{Update Choice}'', with each refinement iteration $q$ further improving the orientation map $\theta^{(q)}(\vc{m})$ according to the cost model. Note that for each pixel $\vc{m}$ of the reference view $\vc{k}_{\rm r}$, the orientation map of each iteration $q$ is obtained by running these two modules in sequence. For refinement iterations $q$ that are even, the pixels of the reference view are processed from left to right and top to bottom, while for refinement iterations $q$ that are odd, those pixels are processed from right to left and bottom to top.

\subsection{Orientation Initialisation}
\label{subsec:orientation-initialization}
An initial orientation map $\theta^{(0)}(\vc{m})$ is required for the IOADR algorithm to converge to an optimal angle map in a reasonable number of iterations. 
Ideally, such initial orientation should be obtained by a low-complexity method.

In this regard, gradient-based approaches are good options, providing, with low computational complexity, good accuracy in non-occluded and non-flat regions~\cite{Wanner2012}. A straightforward implementation of structure-tensor-based depth estimation proved { to be} sufficiently accurate { to obtain} an initial orientation map for the IOADR algorithm. 

The structure tensor is calculated separately for each colour channel of both the horizontal and vertical EPIs of the light field, resulting in six different orientation maps. The initial 4D-PPP orientation map $\theta^{(0)}(\vc{m})$ is obtained by choosing, for each pixel, the orientation $\theta$ corresponding to the disparity value with the highest structure tensor reliability measure, calculated as in~\cite{Bigun1987}.

\subsection{Candidate Orientation Estimation and Evaluation}
\label{subsec:estimate-orientation-candidates-and-costs}
\label{subsubsec:candidate_heuristics}

To improve the initial orientation map $\theta^{(0)}(\vc{m})$, the following steps are executed:
\begin{enumerate}[{Step} 1:]
    \item estimate, for each pixel $\vc{m}$, a set {$\set{C}$} of candidate orientations.
    \item evaluate the cost of each of these candidate orientations { according to Equation \ref{eq:cost}}.
    \item choose the best candidate orientation { as the one that incurs the lowest cost}.
\end{enumerate}

{The success of this process} depends directly on the choice of heuristics used to estimate the set ${\cal C}$ of candidate orientations in Step 1 above.    
{ As the computation of the cost model from Equation~\eqref{eq:cost} is high, it is helpful to devise heuristics to compute a set of candidate orientations { $\set{C} = \{\theta^0_{\rm cnd},\cdots,\theta^N_{\rm cnd}\}$}.  }

Schilling \textit{et al.} \cite{Schilling2018} propose using { orientations from neighbouring samples (referred to here as the {\em Smooth Geometry} heuristic), together with random orientation perturbations as candidates (referred to here as the {\em Random Peturbation} heuristic). The IOADR algorithm proposes to employ, besides those two heuristics, two additional ones: the {\em Colour-Orientation Congruence} heuristic and the {\em Smooth Geometry} heuristic. In what follows, each one of these four heuristics is explained in further detail.}

\begin{enumerate}[a)]
\item \textit{Smooth Depth:} this heuristic indicates that, with { the exception of occlusions, adjacent pixels should belong to the same object and thus have similar depth. This justifies the use of the orientation $\theta(\vc{m})$} of neighbouring samples $\vc{m}$ as candidate orientations.

As in the work of Schilling \textit{et al.}~\cite{Schilling2018}, only the orientations $\theta(\vc{m})$ of the neighbouring samples $\vc{m}$ of the reference view which have already been updated in the current iteration $q$ are added to the candidate set $\set{C}$.

\item \textit{Colour-Orientation Congruence:} this heuristic is based on the Colour-\linebreak Orientation Congruence cost from Subsection~\ref{subsubsubsec:cocc}, which was designed to promote smoothness in the orientation map when the reference view is smooth in terms of colour. {  This cost is computed by comparing the candidate orientation $\theta$ with a smoothed orientation $\theta_{\rm s}(\vc{m}_0)$ (Equation~\eqref{eq:smoothing-term}). 

This smoothed orientation  $\theta_{\rm s}(\vc{m}_0)$ is added to the set ${\cal C}$ of candidate orientations, since it inherently provides optimal results in terms of colour-orientation congruence and requires no additional computations.}

\item \textit{Smooth Plane Geometry:}
{ this heuristic is based on the Planar Geometry cost from Subsection~\ref{subsec:planar-geometry-cost}, which  
requires the computation of an estimated orientation $\theta_\mu$ (see Equation \eqref{eq:pg-cost}). This orientation $\theta_\mu$ would be the correct one in the ideal case where the region is planar. As such, $\theta_\mu$ is added to  the set $\set{C}$ only when the the region is planar, that is, in cases where $\overline{J}_{\rm pg}(\theta,\vc{m}_0,\theta_{\rm map}(\vc{m})) = 0$ as per Equation \eqref{eq:pg-cost}. }

\item \textit{Random Perturbation:} this heuristic follows the procedure suggested in \cite{Schilling2018}. In it, a random small perturbation of the current 4D-PPP orientation is used as a candidate, allowing the algorithm to move away from a sub-optimal result even when the other heuristics fail for a given sample $\vc{m}$ of the reference view of light field.
The orientation  $\theta_{\rm rnd}$ with such a small perturbation can be obtained from:
\begin{equation}
    \theta_{\rm rnd} = \tan^{-1}\left(\tan\theta^{(q-1)}(\vc{m}) + \zeta\right),
\end{equation}
where $\zeta$ is sampled from a normal distribution with zero mean. It has been determined empirically that a standard deviation of $0.04$ provides a good compromise for all light fields.  
\end{enumerate}

\subsection{Update Choice}
\label{subsec:update-choice}
{ At each iteration $q$, the Candidate Orientation Estimation and Evaluation described in Subsection~\ref{subsec:estimate-orientation-candidates-and-costs} provides, for a given sample $\vc{m}_0$ of the reference view $\vc{k}_{\rm r}$, a candidate orientation $\theta_{\rm cnd}$. The IOADR algorithm proposes to include a final step, which makes a stochastic decision between $\theta_{\rm cnd}$ and the previous iteration orientation $\theta^{(q-1)}(\vc{m}_0)$ associated to the sample $\vc{m}_{0}$ of the reference view. 

Such a stochastic decision is performed} using a procedure inspired by the simulated-annealing algorithm~\cite{bertsimas1993}.
In order to do so, a threshold $P$ is computed for each sample $\vc{m}_{0}$ of the reference view from the costs obtained using Equation \eqref{eq:cost} as:
{
\begin{equation}
     P = e^{\frac{J_{\rm old} - J_{\rm cnd}}{T(q)}},
\end{equation}  }
 where $ T(q)$ is the ``temperature'' parameter for the current iteration $q$, { $J_{\rm old}$ is the cost for sample $\vc{m}_{0}$ considering the orientation of the previous iteration $\theta^{(q-1)}$, and $J_{\rm cnd}$ is the cost for the same sample $\vc{m}_{0}$, but considering the orientation $\theta_{\rm cnd}$. If $P \geq 1$, then it means that $J_{\rm cnd}<J_{\rm old}$}, and the candidate orientation $\theta_{\rm cnd}$ should always replace the current estimate. If $P < 1$, the previous iteration orientation $\theta^{(q-1)}(\vc{m}_0)$ will only be replaced by $\theta_{\rm cnd}$ with a probability equal to $P$.
 This decreases the likelihood of the algorithm being trapped at local minima~\cite{bertsimas1993}.

 This process is repeated for all pixels of the reference view of the light field over all iterations $q$. The temperature $T(q)$ is decreased over the iterations $q$ according to an exponential multiplicative cooling schedule, as suggested in~\cite{Kirkpatrick1983}:
\begin{equation}
    T(q) = T_0 \alpha^{\floor{\frac{q}{2}}},
    \label{eq:cooling-schedule}
\end{equation}
with $0<\alpha<1$. The initial ``temperature'' $T_0$ and $\alpha$ are parameters of the algorithm.

\section{Experimental Results}
\label{sec:results}

In this section the performance of the proposed IOADR algorithm is assessed. It is divided into three parts. The first describes the experimental conditions. The second shows a comparison between results achieved with the proposed method and the state-of-the-art (SOTA). The third presents Ablation Studies, where each of the contributions of the proposed framework is analysed individually, by evaluating the effects of excluding a specific contribution. 

\subsection{Experimental Conditions}
\begin{figure}[ht]
    \centering
    
    \subfloat{\includegraphics[width=0.10\textwidth]{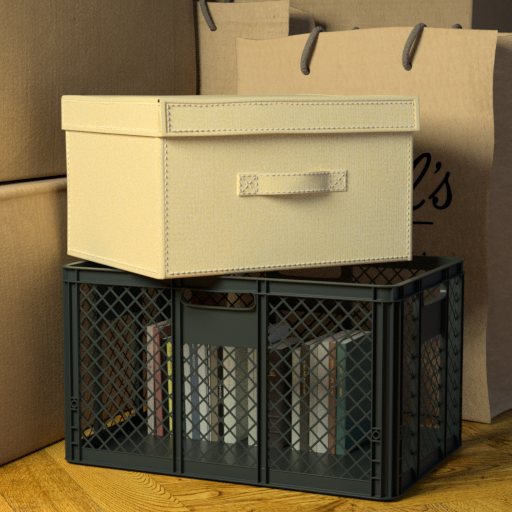}%
    \hfil
    \includegraphics[width=0.10\textwidth]{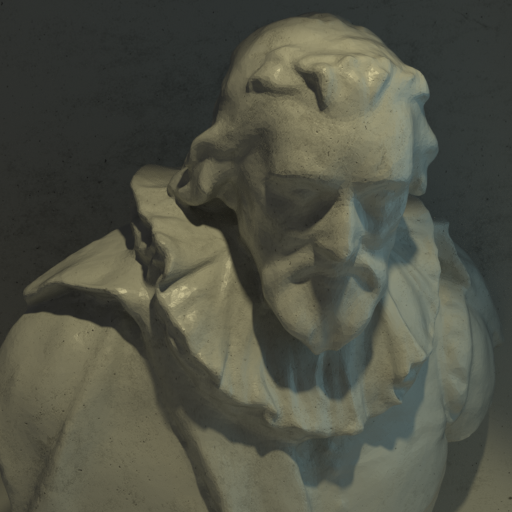}%
    \hfil
    \includegraphics[width=0.10\textwidth]{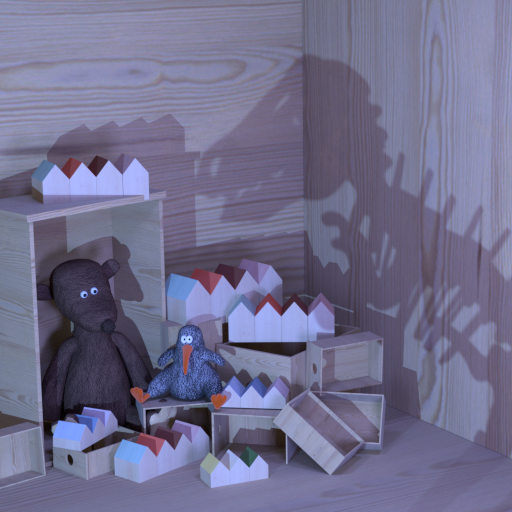}%
    \hfil
    \includegraphics[width=0.10\textwidth]{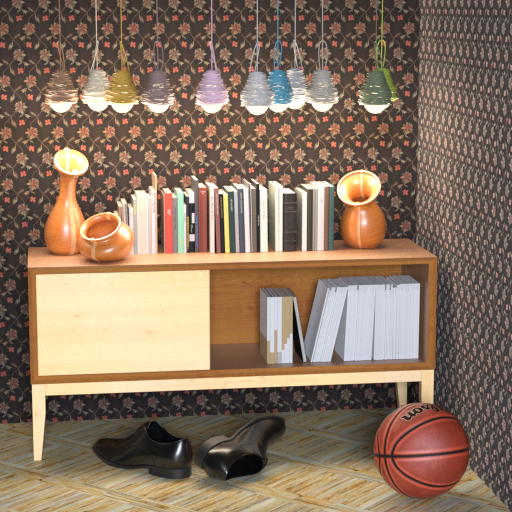}}\\
    \subfloat{\includegraphics[width=0.10\textwidth]{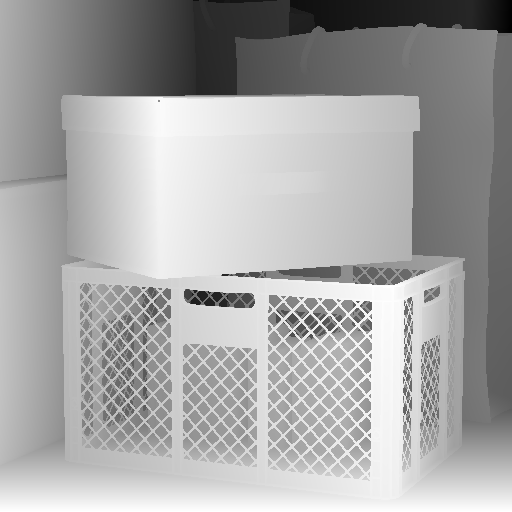}\hfil
    \includegraphics[width=0.10\textwidth]{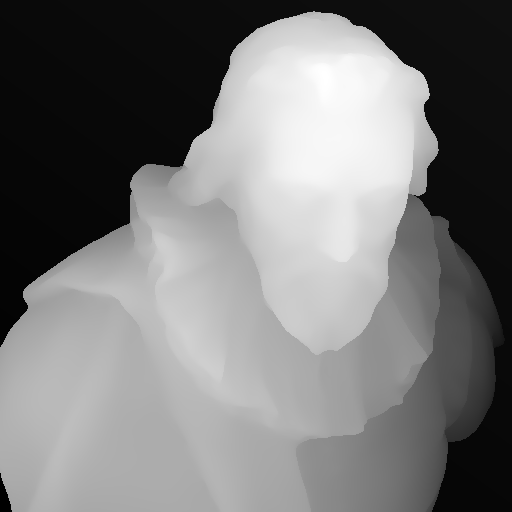}\hfil
    \includegraphics[width=0.10\textwidth]{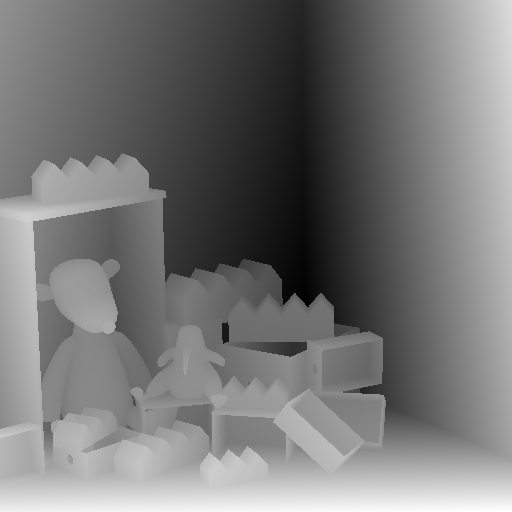}\hfil%
    \includegraphics[width=0.10\textwidth]{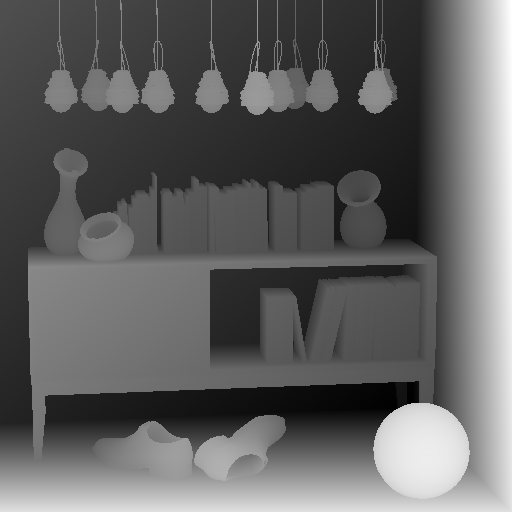}}\\
    
    \caption{Centre views and respective ground truth disparity maps of the (from left to right) Boxes, Cotton, Dino, and Sideboard light fields from the HCI 4D Lightfield dataset~\cite{Johansen2017}.}
    \label{fig:dataset}
\end{figure}

The IOADR algorithm was implemented in C++ and its source code is available at \href{https://github.com/RuiLourenco/IOADR}{https://github.com/RuiLourenco/IOADR}, { where experiments with additional light field datasets and further visual comparisons are available}. The algorithm is tested using the HCI 4D Lightfield dataset~\cite{Johansen2017}, a computer-generated light field dataset that provides ground truth disparity maps, which allow for objective comparisons. The centre view of each light field and the respective ground truth disparity are shown in Figure~\ref{fig:dataset}. All light fields in the HCI 4D Lightfield dataset~\cite{Johansen2017} have equal disparities in the $s\times u$ and $t\times v$ planes. Referring to Equation~\eqref{eq:disparity}, this is equivalent to having $\vc{\eta} = \nmatrix{\eta,\eta}$.

The assessment of the depth estimation techniques is performed using three metrics: two evaluate pixel-wise accuracy -- the MSE $\times 100$ and the badpix 0.07; the third one -- the MAE in planar regions -- evaluates the error in the estimation of surface normals calculated from the depths (Section \ref{subsec:known-limitations}), which takes into account sets of neighbouring depths. This metric is calculated exclusively in planar regions of the light fields using a ground-truth binary map provided along with the dataset. These three objective metrics are defined in~\cite{Johansen2017}.

The IOADR algorithm includes fourteen tunable parameters;  in the experimental results shown in the sequel, their values are as shown in Table~\ref{tab:variables}. While tuning these parameters separately for each light field provides a small increase in performance, in this paper, these parameters are the ones shown in Table~\ref{tab:variables} for all light fields.  

\begin{table}[ht]
    \centering
    \caption{Values for different parameters used in the proposed algorithm. $\eta$ is the value that converts the orientation of the 4D-PPP into a disparity (Equation~\eqref{eq:disparity}).}
     \label{tab:variables}

    \begin{tabular}{ccccccc}
        \toprule
         $T_0$ & $q_{\rm max}$& $\alpha$&$\epsilon_\theta$& $\rho_{c}$ &$\rho_{\theta}$&$\tau_{\rm c}$\\\midrule
         10& 10 &0.8 &0.5 &0.15 &10&3 \\
         \bottomrule
    \end{tabular}
        \vspace{0.5em}

        \begin{tabular}{cccccccc}
        \toprule
         $\tau_{\epsilon}$&$\tau_\theta$ &$\tau_{\rm a}$&$\delta_a$&$\sigma_{\rm a}$&$\lambda_0$ &$\gamma_0$\\\midrule
         $\frac{0.031}{\eta}$&$\frac{0.031}{\eta}$ & 1.3&5&0.04&100&0.05\\
         \bottomrule
    \end{tabular}
    \vspace{-2em}
\end{table}

\subsection{Comparison with State-of-the-Art}

Table \ref{tab:state-of-the-art} compares the proposed IOADR algorithm with five different SOTA algorithms. Three traditional non-learning based (Ober-Cross~\cite{Schilling2018}, Ober-Cross + ANP~\cite{Schilling2018}, and OFSY~\cite{Strecke2018}), { one unsupervised learning-based method (DispNet+OccNet~\cite{Zhang2023}) and three learning-based methods---SubFocal-L~\cite{Wentao2023}, OACC-Net~\cite{Wang2022} and} AttNet~\cite{Tsai2020}. The comparison is made in terms of the metrics MSE $\times 100$, Badpix 0.07, and (when available) MAE in planar regions, as defined in~\cite{Johansen2017}. 

\begin{table}[H]
  \centering
  \caption{Objective Metric Comparison with State-of-the-Art methods. Shaded cells indicate supervised methods. The best result for each light field is in bold. The best non-learning-based result for each light field is underlined.}
    \begin{tabular}{lrrrr}
    \toprule
          & \multicolumn{1}{l}{Boxes} & \multicolumn{1}{l}{Cotton} & \multicolumn{1}{l}{Dino} & \multicolumn{1}{l}{Sideboard} \\
    \midrule
          & \multicolumn{4}{c}{MSE $\times$100} \\
    \cmidrule{2-5}
        \rowcolor{Gray}
    SubFocal-L\cite{Wentao2023} &\textbf{2.417} &0.243 &0.1013&0.441 \\
    \rowcolor{Gray}
    OACC-Net\cite{Wang2022} &2.892 &0.162 &0.083&0.542 \\
\rowcolor{Gray}
AttNet\cite{Tsai2020}         & 3.842 & \textbf{0.059} & \textbf{0.045} & \textbf{0.398} \\
    DispNet+OccNet\cite{Zhang2023} & NA & NA & 0.650 & 1.738\\
    Ober-Cross + ANP\cite{Schilling2018} & 4.750 & 0.555 & 0.336 & \underline{0.941} \\
    Ober-Cross\cite{Schilling2018} & \underline{4.160} & 0.501 & \underline{0.309} & 0.963 \\
    OFSY\cite{Strecke2018}   & 9.561 & 2.653 & 0.782 & 2.478 \\
    IOADR (Ours)  & 4.601 & \underline{0.375} & 0.319 & 0.962 \\
    \midrule
          & \multicolumn{4}{c}{Badpix 0.07} \\
\cmidrule{2-5}
\rowcolor{Gray} SubFocal-L\cite{Wentao2023}  &\textbf{7.27}\%                   & 0.25\%          & 0.68\%          & \textbf{2.69\%}          \\
\rowcolor{Gray} OACC-Net\cite{Wang2022} &10.70\%                   & 0.31\%          & 0.97\%          & 3.35\%          \\
\rowcolor{Gray}
AttNet\cite{Tsai2020}         &11.14\%                   & \textbf{0.20\%} & \textbf{0.44\%} & \textbf{2.69\%} \\

    DispNet+OccNet\cite{Zhang2023} & NA & NA & 6.59\% & 12.01\% \\
    Ober-Cross + ANP\cite{Schilling2018} & \underline{10.76\%} & 1.02\% & 2.07\% & \underline{5.67\%} \\
    Ober-Cross\cite{Schilling2018} & 13.13\% & \underline{0.94\%} & \underline{1.95\%} & 6.28\% \\
    OFSY\cite{Strecke2018}  & 19.25\% & 3.04\% & 3.43\% & 10.36\% \\
    IOADR (Ours)   & 15.53\% & 2.21\% & 3.09\% & 8.01\% \\
    \midrule
          & \multicolumn{4}{c}{MAE in planar regions} \\
\cmidrule{2-5} 
\rowcolor{Gray} SubFocal-L\cite{Wentao2023}  &4.702 & 13.112 & 4.904 & 5.158 \\
\rowcolor{Gray} OACCNet\cite{Wang2022} &7.004 & 21.584 & 6.014 & 6.337 \\
\rowcolor{Gray} AttNet\cite{Tsai2020} & 5.819 & 10.472 & 2.686 & 6.078 \\
Ober-Cross + ANP\cite{Schilling2018} & 5.402 & 12.674 & 3.062 & 6.219 \\
    Ober-Cross\cite{Schilling2018}  & 8.894 & 15.951 & 4.893 & 12.083 \\
    OFSY\cite{Strecke2018}  & 3.574 & 2.909 & 1.069 & 4.151 \\
    IOADR (Ours)   & \underline{\textbf{1.819}} & \underline{\textbf{2.885}} & \underline{\textbf{0.593}} & \underline{\textbf{3.706}} \\
    \bottomrule
    \end{tabular}%
  \label{tab:state-of-the-art}%
  \vspace{-1.5em}
\end{table}%

The proposed IOADR algorithm proves to be superior to all SOTA methods in terms of MAE in Planar Regions, achieving, on average, a 26.3\% better result when compared to the second-best method, OFSY \cite{Strecke2018}, which explicitly focuses on the accuracy of surface normals. Furthermore, the proposed algorithm outperforms the OFSY algorithm in terms of MSE$\times100$ and Badpix 0.07 for all assessed light fields, { showing that the improved geometric accuracy does not come at the expense of pixel-wise accuracy.}

Regarding pixel-wise accuracy, as measured by MSE$\times100$ and Badpix 0.07, the IOADR algorithm { is competitive in terms of both pixel-wise accuracy metrics when compared to the unsupervised algorithms, although it underperforms relative to the supervised algorithms.

This provides evidence that the geometrical concepts presented in Section \ref{sec:4dppp}, together with the costs and heuristics proposed based on these geometrical insights, open a promising avenue of research that may lead to further improvements in both traditional and supervised learning-based methods.}

\subsection{Ablation Studies}
This section assesses the effectiveness of the costs and heuristics proposed in this paper, namely the Occlusion Awareness, the multi-term energy cost model, and the various heuristics to estimate candidate orientations $\theta$. The baseline is given by the IOADR results shown in Table~\ref{tab:state-of-the-art}.

\paragraph{Occlusion Awareness} ~ \\
{ in this section, we assess the effectiveness of the} Occlusion-Aware Pixel Deviation (OAPD) used as the novel data cost $J_{\rm oa}$ in Equation~\eqref{eq:data-cost}, { relative to using}  the simple Pixel Deviation as in Equation~\eqref{eq:pixel-deviation}. 


\begin{table}[t]
  \centering
  \caption{Ablation studies assessing MSE$\times$100 and Badpix 0.07 results with and without the occlusion aware data cost $J_{\rm oa}$. The best results for each light field are in bold.}
    \begin{tabular}{lcccc}
    \toprule
    \multicolumn{1}{l}{Data Cost Term} & \multicolumn{1}{l}{Boxes} & \multicolumn{1}{l}{Cotton} & \multicolumn{1}{l}{Dino} & \multicolumn{1}{l}{Sideboard} \\
    \midrule
          & \multicolumn{4}{c}{MSE $\times$100} \\\cmidrule{2-5}
    $J_{\rm pd}$ (Equation~\eqref{eq:pixel-deviation})    & 9.472&	4.113&0.832&3.104\\
    $J_{\rm oa}$ (Equation~\eqref{eq:data-cost}) &\textbf{4.601}&\textbf{0.375}&\textbf{0.319}&\textbf{0.962}\\
    \midrule
          & \multicolumn{4}{c}{Badpix 0.07} \\\cmidrule{2-5}
     $J_{\rm pd}$ (Equation~\eqref{eq:pixel-deviation})   & 19.94\%&	5.28\%&	5.88\%&	13.07\%
 \\
    $J_{\rm oa}$ (Equation~\eqref{eq:data-cost})  &\textbf{15.53}\%&\textbf{2.21}\%&\textbf{3.09}\%&\textbf{8.01}\%\\\bottomrule
    \end{tabular}%
  \label{tab:oa-performance}%
\end{table}%

Table \ref{tab:oa-performance} shows the MSE$\times$100 and Badpix 0.07 results for computer-generated light fields. When compared with the use of the simple Pixel Deviation $J_{\rm pd}$, the use of the proposed occlusion aware data cost $J_{\rm oa}$ provides a steep reduction in both metrics for all four tested light fields, achieving, for example, a 90\% reduction in MSE$\times$100 for the Cotton light field.

\begin{figure}[t]
    \centering
    
    \subfloat[]{\includegraphics[width=0.21\columnwidth]{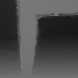}\label{}}%
    \hfil
    \subfloat[]{\includegraphics[width=0.21\columnwidth]{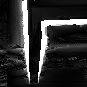} \label{}}%
     \subfloat[]{\includegraphics[width=0.21\columnwidth]{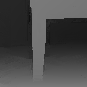}\label{}}%
    \hfil
    \subfloat[]{\includegraphics[width=0.21\columnwidth]{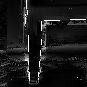} \label{}}%
    
    \caption{Occlusion detail for the light field \textit{Sideboard}: (a) and (c) show disparity maps using, respectively,  the simple Pixel Deviation $J_{\rm pd}$ in Equation~\eqref{eq:pixel-deviation} and the proposed data cost $J_{\rm oa}$ in Equation~\eqref{eq:data-cost}; (b) and (d) show the absolute difference between the disparities in (a) and (c), respectively, and the ground truth. White corresponds to the largest error, and black corresponds to no error.}
    \label{fig:oa-performance-sideboard}  
    \vspace{-1em}
\end{figure}

This is exemplified in Figure \ref{fig:oa-performance-sideboard}, which shows a visual comparison of the disparity maps obtained, for the light field \textit{Sideboard}, by the IOADR algorithm using either the proposed Occlusion-Aware data cost $J_{\rm oa}$ from Equation~\eqref{eq:data-cost} or the simple Pixel Deviation $J_{\rm pd}$ from Equation~\eqref{eq:pixel-deviation}. 
The borders are far more accurate when $J_{\rm oa}$ is used instead  $J_{\rm pd}$, which shows the greater effectiveness of the proposed Occlusion-Aware Data Cost when compared with the simple Pixel Deviation. 

\paragraph{Energy Cost Model} ~ \\
the three-factor cost model, described in Equation~\eqref{eq:cost}, is another contribution of this work. The impact of the inclusion of the Colour-Orientation Congruence term $J_{\rm coc}$ and of the Plane Geometry term $J_{\rm pg}$ in the data cost is shown in Table \ref{tab:cost-composition}. It compares the MSE $\times 100$, Bad Pix, and MAE in planar Regions obtained when using the full cost in Equation~\eqref{eq:cost}, to the ones obtained when excluding both the $J_{\rm pg}$ and $J_{\rm coc}$ terms, and when excluding only either $J_{\rm pg}$ or $J_{\rm coc}$. As can be seen in  Table \ref{tab:cost-composition}, the use of $J_{\rm coc}$ significantly improves all three metrics. The use of $J_{\rm pg}$ also leads to sizeable improvements in the MAE in planar regions, but at the expense of slightly higher  MSE$\times$100 and badpix 0.7 metrics.

\begin{table}[t]
  \centering
  \caption{Ablation studies assessing MSE$\times$100, Badpix 0.07 and MAE when excluding some of the different terms of the total energy cost $J$ in Equation~\eqref{eq:cost}. The best result for each light field is in bold.}
    \begin{tabular}{ccccccc}
    \toprule
     \multicolumn{3}{c}{Cost Terms (Equation~\eqref{eq:cost})} & \multicolumn{1}{l}{Boxes} & \multicolumn{1}{l}{Cotton} & \multicolumn{1}{l}{Dino} & \multicolumn{1}{l}{Sideboard} \\
    \cmidrule(lr){1-3}\cmidrule(lr){4-7}
    $J_{\rm pg}$ & $J_{\rm coc}$ & $J_{\rm oa} $ & \multicolumn{4}{c}{MSE$\times$100} \\
    \cmidrule(lr){1-3} \cmidrule(lr){4-7}
      & X & X & \textbf{4.590}	&\textbf{0.356	}&\textbf{0.312	}&\textbf{0.962} \\
    X &   & X &5.680&1.202&	0.936	&1.818\\
      &   & X & 5.490	&1.006	&0.766	&1.717\\
    X & X & X & 4.601&0.375&0.319&\textbf{0.962}
\\
    \midrule
    $J_{\rm pg}$ & $J_{\rm coc}$ & $J_{\rm oa} $ & \multicolumn{4}{c}{Badpix 0.07} \\
    \cmidrule(lr){1-3} \cmidrule(lr){4-7}
      & X & X &\textbf{15.27}\%	&\textbf{1.59}\%	&\textbf{2.85}\%	&\textbf{7.86}\%
 \\
    X  &   & X & 34.07\%&21.27\%&18.26\%&22.41\%
 \\
      &   & X & 27.21\%&	12.07\%&11.90\%&18.56\%
\\
    X & X & X & 15.53\%&2.21\%&3.09\%&8.01\%\\
    \midrule
    $J_{\rm pg}$ & $J_{\rm coc}$ & $J_{\rm oa} $ & \multicolumn{4}{c}{MAE  in planar regions} \\
    \cmidrule(lr){1-3} \cmidrule(lr){4-7}
      & X & X & 6.327&	3.651&	5.188	&8.889
 \\
    X &   & X & 64.317&82.951&	28.962&50.214\\
      &   & X & 52.748&79.040&23.248&47.952 \\
    X & X & X &\textbf{1.819}&\textbf{2.885}&\textbf{0.593}&\textbf{3.706}
 \\
    \bottomrule
    \end{tabular}%
  \label{tab:cost-composition}%
  \vspace{-0.5em}
\end{table}%

\paragraph{Candidate Orientations} ~ \\
{ the performance changes deriving from the exclusion of each candidate orientation heuristic proposed in} Subsection~\ref{subsubsec:candidate_heuristics} is also assessed in terms of MSE$\times$100, Badpix 0.07, and MAE in planar regions.

\begin{table}[t]
  \centering
  \caption{Ablation studies assessing MSE$\times$100, Badpix 0.07, and MAE in planar regions when excluding each one of the candidate orientation estimation heuristics defined in Section \ref{subsec:estimate-orientation-candidates-and-costs} from the framework. The best result for each light field is in bold.}
    \begin{tabular}{lcccc}
    \toprule
     & \multicolumn{1}{l}{Boxes} & \multicolumn{1}{l}{Cotton} & \multicolumn{1}{l}{Dino} & \multicolumn{1}{l}{Sideboard} \\
    \cmidrule(lr){2-5}
     Excluded Heuristics & \multicolumn{4}{c}{MSE $\times$100} \\
    \cmidrule(lr){1-1} \cmidrule(lr){2-5}
    Smooth Depth & 9.401	&2.686&	0.640&	1.566\\
    C-O Congruence &4.692	&0.376	&0.345	&1.004\\
    Smooth Plane &4.768	&0.538&	0.479&	1.128 \\
    None & \textbf{4.601}&\textbf{0.375}&\textbf{0.319}&\textbf{0.962}
 \\
    \midrule
    Excluded Heuristics & \multicolumn{4}{c}{Badpix 0.07} \\
    \cmidrule(lr){1-1} \cmidrule(lr){2-5}   
    Smooth Depth  & 24.51\%	&7.21\%	&7.79\%	&13.15\%
 \\
    C-O Congruence & \textbf{14.77}\%&	\textbf{2.17}\%&	3.34\%&	8.24\%
\\
    Smooth Plane & 26.08\%&	14.71\%&	14.47\%&	18.21\%
 \\
    None  &15.53\%&2.21\%&\textbf{3.09}\%&\textbf{8.01}\%
 \\\midrule
    Excluded Heuristics  & \multicolumn{4}{c}{MAE in planar regions} \\
    \cmidrule(lr){1-1} \cmidrule(lr){2-5}    
    Smooth Depth  & 2.268	&9.253	&1.038&4.044
 \\
    C-O Congruence & 2.390	&2.913	&0.716	&4.424 \\
    Smooth Plane & 56.882&	81.628	&50.197&	56.153 \\
    None  & \textbf{1.819} & \textbf{2.885} & \textbf{0.593} & \textbf{3.706} \\
    \bottomrule
    \end{tabular}%
    
  \label{tab:candidate-heuristics}%
  \vspace{-1.5em}
\end{table}%

Table \ref{tab:candidate-heuristics} shows the results achieved with the proposed algorithm when the orientation candidates are determined by foregoing each one of the candidate orientation heuristics:  Smooth Depth, Colour-Orientation Congruence (C-O Congruence), and Smooth Plane heuristics. On one hand, the significant loss in accuracy obtained from forgoing the Smooth Depth heuristic and the Smooth Plane Heuristic is worth noticing, both demonstrating significant increases in both MSE$\times$100 and Badpix 0.07 when excluded. Foregoing the Smooth Plane Heuristic additionally results in a major increase in MAE in planar regions.

On the other hand, the inclusion of candidate orientations based on Colour-Orientation Congruence has a lesser impact on the accuracy of the resultant disparity map in terms of MSE$\times$100 and Badpix 0.07; in fact, not including these candidates even produces minor improvements in the Badpix 0.07 metric for some light fields. However, its use leads to a significant improvement in terms of MAE in planar regions. 

\section{Conclusion}
\label{sec:conclusion}
This paper introduces a formal mathematical framework for describing depth estimation based on 4D light field geometry. This framework was shown to be helpful in analysing and addressing the limitations of 4D light field depth estimation. Based on this study,  a novel light field depth estimation algorithm (IOADR) is proposed, based on a local optimisation method for depth estimation with three significant contributions: a novel occlusion detection algorithm capable of delivering occlusion-aware disparity estimation with accurate boundaries, a new algorithm for estimating accurate surface normals from noisy depth estimations, and a cost term capable of evaluating for each depth value candidate its suitability to be a good fit relative to the local surface normals.

The proposed IOADR algorithm presents competitive results regarding MSE$\times$100 when compared to other non-learning-based methods. In terms of MAE in planar regions, IOADR achieves, by far, the best results when compared to both non-learning-based and learning-based methods;  such good results in the planar areas are obtained without overly compromising the accuracy of the disparity map. For the MAE in planar regions metric, the proposed method obtains, on average, a gain of $26.3\%$ relative to the second-best method.

It is important to note that the mathematical framework formalised in this paper may bring valuable insights for the development of both learning and non-learning-based methods. For instance, the proposed cost model in Equation~\eqref{eq:cost} may be readily incorporated into a learning-based architecture, which opens a promising research avenue.

\bibliographystyle{IEEEtran}
\bibliography{citations}

%

%








\end{document}